\documentclass{jfm}
\usepackage{afterpage}
\usepackage{longtable}
\usepackage[english]{babel}
\usepackage{dcolumn}
\usepackage{bm}
\usepackage{pifont}
\usepackage{amsfonts}
\usepackage{amsmath}
\usepackage{upmath}
\usepackage{amssymb}
\usepackage{wasysym}
\usepackage{amsbsy}
\usepackage{graphicx}
\usepackage{natbib}
\usepackage{psfrag}
\usepackage{lineno}
\usepackage{color}
\usepackage{multirow}
\usepackage{cases}
\usepackage{stackengine}
\usepackage{float}
\usepackage{blkarray}

\newcommand*\patchAmsMathEnvironmentForLineno[1]{%
  \expandafter\let\csname old#1\expandafter\endcsname\csname #1\endcsname
  \expandafter\let\csname oldend#1\expandafter\endcsname\csname end#1\endcsname
  \renewenvironment{#1}%
     {\linenomath\csname old#1\endcsname}%
     {\csname oldend#1\endcsname\endlinenomath}}%
\newcommand*\patchBothAmsMathEnvironmentsForLineno[1]{%
  \patchAmsMathEnvironmentForLineno{#1}%
  \patchAmsMathEnvironmentForLineno{#1*}}%
\AtBeginDocument{%
\patchBothAmsMathEnvironmentsForLineno{equation}%
\patchBothAmsMathEnvironmentsForLineno{align}%
\patchBothAmsMathEnvironmentsForLineno{flalign}%
\patchBothAmsMathEnvironmentsForLineno{alignat}%
\patchBothAmsMathEnvironmentsForLineno{gather}%
\patchBothAmsMathEnvironmentsForLineno{multline}%
}

\ifCUPmtlplainloaded \else
  \checkfont{eurm10}
  \iffontfound
    \IfFileExists{upmath.sty}
      {\typeout{^^JFound AMS Euler Roman fonts on the system,
                   using the 'upmath' package.^^J}%
       \usepackage{upmath}}
      {\typeout{^^JFound AMS Euler Roman fonts on the system, but you
                   dont seem to have the}%
       \typeout{'upmath' package installed. JFM.cls can take advantage
                 of these fonts,^^Jif you use 'upmath' package.^^J}%
      }
  \else
  \fi
\fi

\ifCUPmtlplainloaded \else
  \checkfont{msam10}
  \iffontfound
    \IfFileExists{amssymb.sty}
      {\typeout{^^JFound AMS Symbol fonts on the system, using the
                'amssymb' package.^^J}%
       \usepackage{amssymb}%
         \let\leq=\leqslant
         
      }{}
  \fi
\fi

\ifCUPmtlplainloaded \else
  \IfFileExists{amsbsy.sty}
    {\typeout{^^JFound the 'amsbsy' package on the system, using it.^^J}%
     \usepackage{amsbsy}}
    {\providecommand\boldsymbol[1]{\mbox{\boldmath $##1$}}}
\fi

\newcommand{\D}{\ensuremath{\mathrm{d}}}

\newcommand{\ms}{\kern.10em\relax}
\newcommand{\Web}{\mbox{\textit{We}}} 		% Weber
\newcommand{\Bo}{\mbox{\textit{Bo}}} 		% Bond
\newcommand{\Ka}{\mbox{$\Gamma$}} 		% Kapitza

\providecommand\bcdot{\boldsymbol{\cdot}}
\newcommand{\boldm}[1]{\boldsymbol{#1}}

\newcommand{\eg}{e.g.\ }

\title[The nonlinear states of viscous capillary jets confined in the axial direction]%
{The nonlinear states of viscous capillary jets confined in the axial direction}

\author%
[A. Mart\'inez-Calvo, M. Rubio-Rubio, A. Sevilla]%
{A. Mart\'inez-Calvo$^1$,\ns
M. Rubio-Rubio$^2$,\ns
and A. Sevilla$^1$\thanks{Email address for correspondence: alejandro.sevilla@uc3m.es}}

\affiliation{%
$^1$Grupo de Mec\'anica de Fluidos,
Departamento de Ingenier\'ia T\'ermica y de Fluidos,
Universidad Carlos III de Madrid,
Av.~Universidad 30,
28911 Legan\'es (Madrid),
Spain\\[\affilskip]
$^2$\'Area de Mec\'anica de Fluidos,
Departamento de Ingenier\'ia Mec\'anica y Minera, 
Universidad de Ja\'en,
Campus de las Lagunillas,
23071, Ja\'en (Spain)}

\pubyear{2017}
\begin{document}
\maketitle
%\linenumbers

\begin{abstract}
We report an experimental and theoretical study of the global stability and nonlinear dynamics of vertical jets of viscous liquid confined in the axial direction due to their impact on a bath of the same liquid. Previous works demonstrated that in the absence of axial confinement the steady liquid thread becomes unstable due to an axisymmetric global mode for values of the flow rate, $Q$, below a certain critical value, $Q_c$, giving rise to oscillations of increasing amplitude that finally lead to a dripping regime~\citep{SauteryBuggisch, Mariano}. Here we focus on the effect of the jet length, $L$, on the transitions that take place for decreasing values of $Q$. The linear stability analysis shows good agreement with our experiments, revealing that $Q_c$ increases monotonically with $L$, reaching the semi-infinite jet asymptote for sufficiently large values of $L$. Moreover, as $L$ decreases a quasi-static limit is reached, whereby $Q_c\to 0$ and the neutral conditions are given by a critical length determined by hydrostatics. Our experiments have also revealed the existence of a new regime intermediate between steady jetting and dripping, in which the thread reaches a limit-cycle state without breakup. We thus show that there exist three possible states depending on the values of the control parameters, namely steady jetting, oscillatory jetting and dripping. For two different combinations of liquid viscosity, $\nu$, and injector radius, $R$, the boundaries separating these regimes have been determined in the $Q$-$L$ parameter plane, showing that steady jetting exists for small enough values of $L$ or large enough values of $Q$, dripping prevails for small enough values of $Q$ or sufficiently large values of $L$, and oscillatory jetting takes place in an intermediate region whose size increases with $\nu$ and decreases with $R$.
\end{abstract}

\begin{keywords}
Capillary flows, Absolute/convective instability, Nonlinear instability.
\end{keywords}

\section{Introduction \label{sec:intro}}

The great research effort devoted in the past to understand and control the dynamics of liquid jets is justified by their rich phenomenology and the large number of technological applications where they play a central role, such as fuel atomization, chemical reactors, ink-jet and 3D printing, additive manufacturing, microfluidic platforms, drug encapsulation, mass spectrometry or cytometry, to name a few~\citep[see \eg the reviews by][]{Bogy79,Eggers97,Lin98,Basaran2002,BarreroAR,Christopher_Anna07,EGG08,Derby10,Anna16}. These applications often require the generation of micrometre-sized jets and drops, in which case it proves convenient to downscale the disperse phase from a typically millimetre-sized injector to the desired micrometer scales to avoid clogging issues. This stretching effect can be achieved through different techniques like fibre spinning~\citep{MatovichPearson,PearsonMatovich}, electrospinning~\citep{Doshi_Reneker_95}, flow focusing~\citep{Ganan98}, viscous co-flows~\citep{Suryo06,Marin09,Evangelio16} and gravitational stretching~\citep{SauteryBuggisch, Mariano}. The latter configuration is particularly simple to implement, and the present work naturally extends the investigation of~\citet{Mariano} to assess the influence of a finite jet length and nonlinearity on the dynamics of the liquid thread.

In the absence of axial confinement, previous studies of the downwards injection of a constant flow rate of liquid into a passive gaseous atmosphere have demonstrated the \mbox{existence} of two different flow states, namely dripping and jetting, respectively prevailing for small and large enough values of the flow rate~\citep{Clanet1999,Ambravaneswaran2004}. The jetting regime is characterised by the formation of a liquid column which breaks-up into drops at a certain distance from the injector due to the downstream growth of axisymmetric capillary instability waves~\citep[see \eg][]{Plateau,Rayleigh1,DonnellyGlaberson,Kalaaji2003,GonzalezGarcia2009}. In contrast, the dripping regime features the emission of comparatively larger drops near the injector~\citep{Wilkes1999,Coullet,Subramani06}. From the point of view of local stability theory, the jetting regime is a convectively unstable flow, in which the downstream advection of growing disturbances by the underlying base flow allows the formation of a long liquid column. Moreover, several works have successfully described the transition from jetting to dripping as a global instability that, in the case of quasi-parallel jets, is linked with the onset of local absolute instability~\citep{LeibyGoldsteinAC,LeibyGoldstein,ledizes97,Vihinen1997,Odonnell2001,Soderberg03,Sevilla2011,Josefa2016}.

The concepts of convective and absolute instability rely on the assumption of quasi-parallel flow, and do not apply to cases where the wavelength of disturbances is of the order of the development length of the base flow, as happens in highly stretched jets like those studied in the present work. In these cases a global stability analysis must be performed, in which the spatial structure of the eigenfunctions is obtained as part of the solution~\citep{Theofilis2011}. The global stability analysis is greatly simplified by the use of one-dimensional approximations to the full conservation equations, since the eigenvalue problem involves only the axial coordinate as an eigendirection~\citep[see \eg][]{PearsonMatovich,SauteryBuggisch,Mariano,gordillo14}. In particular, a global stability analysis of the leading-order one-dimensional model for viscous liquid columns~\citep{GyC,EggersDupont} was first applied to gravitationally stretched viscous jets by~\citet{SauteryBuggisch}, and refined later on by~\citet{Mariano}. The latter works revealed that the axisymmetric self-excited oscillations observed in long viscous jets below a certain critical flow rate are explained by the destabilisation of a linear global mode.

Despite the usefulness of linearised theory to predict many features of liquid jets, there are relevant aspects of their dynamics that can only be described using a nonlinear approach, prominent examples being the pinch-off singularity~\citep{Eggers1993} and the formation of satellite droplets~\citep{Yuen1968, Nayfeh1970, RutlandJameson,ashgriz_mashayek_1995}. In the present work we introduce a new configuration where nonlinearity provides the selection mechanism between two different regimes after the jet becomes globally unstable: either a limit-cycle state without breakup described here for the first time (see movie 2 of the supplementary material), or a fully developed dripping state (see movie 3 of the supplementary material).

Liquid threads of finite length, either by their impingement onto a bath of the same liquid, or by their impact on a solid plate, have also been considered in previous works. Thus, the shape and stability of quasi-static axisymmetric liquid bridges formed between a solid rod and an infinite bath were studied by~\citet{Kovitz75}, and more recently by~\citet{Benilov10} and~\citet{Benilov13}. \citet{Christodoulides2010} studied the steady structure of planar vertical inviscid jets impinging onto a horizontal solid plate. However, most of the literature deals with the phenomenon of coiling, the fascinating buckling instability associated with the impact of a jet or film of viscous liquid on a solid substrate~\citep[see][and references therein]{RibeARFM2012}. Another beautiful and surprisingly rich phenomenon that has been studied is the deposition of viscous threads on moving solid substrates~\citep{chiu-webster_lister_2006,blount_lister_2011}. These coiling states and deposition patterns break the axial symmetry, and their mathematical description is complicated by the fact that the thread centreline must be obtained as part of the solution~\citep{entov_yarin_1984}. Although in the experiments reported herein we have observed coiling under certain conditions, our focus is on the global self-excited oscillations caused by the destabilisation of the axisymmetric breathing mode studied by~\citet{SauteryBuggisch} and~\citet{Mariano}. It should be pointed out that the breathing mode and coiling coexist in a wide region of parameter space, as evidenced by movie 5 of the supplementary material. Nevertheless, in all the cases considered herein the coupling between both modes is weak enough for a purely axisymmetric model to provide a reasonably good leading-order description, not only of the neutral conditions for the onset of the breathing mode, but also of the long-time regime of the jet.

In the present work we report experiments performed to characterize the linear and nonlinear stability properties of jets of Newtonian liquid, injected at a constant flow rate through a circular tube, that impinge on the free surface of a reservoir of the same liquid placed at a controlled distance from the injector. The experiments are complemented with a global stability analysis and numerical simulations, both based on the leading-order one-dimensional mass and momentum equations retaining the full expression of the interfacial curvature~\citep{EggersDupont,Mariano}. We also study the selection of nonlinear regimes under globally unstable conditions. The experimental setup and the mathematical model are described in \S\ref{sec:flow}, and the results are presented in \S\ref{sec:results}, finally leading to the conclusions drawn in~\S\ref{sec:conclusions}.

\section{Experimental setup and mathematical model \label{sec:flow}}

\subsection{Flow configuration and experimental setup \label{subsec:setup}}

\begin{figure}
    \centering
    \includegraphics[width=180pt]{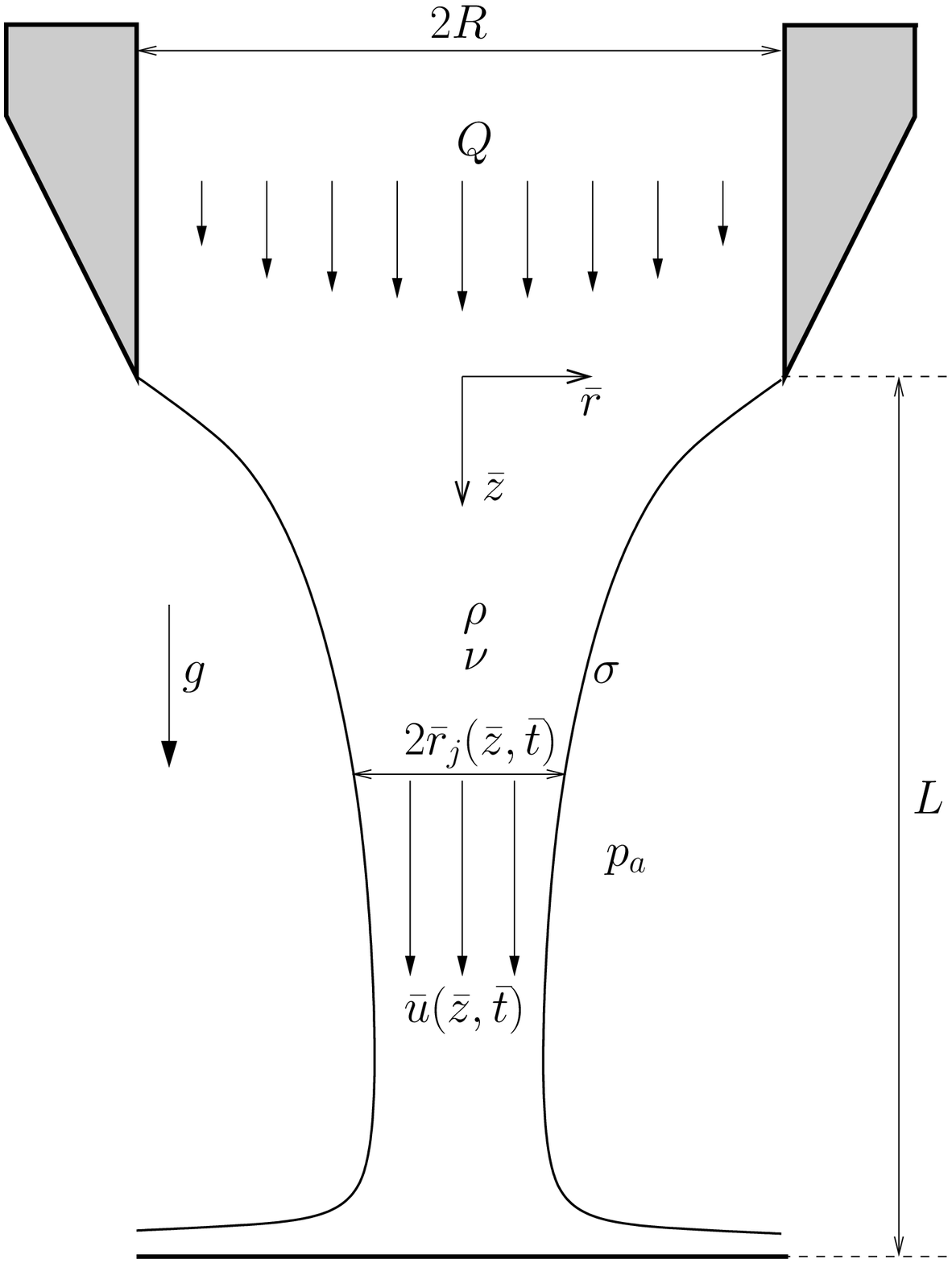}
    \llap{\parbox[b]{200pt}{(\textit{a})\\[215pt]}}
    \hfill
    \includegraphics[width=62pt]{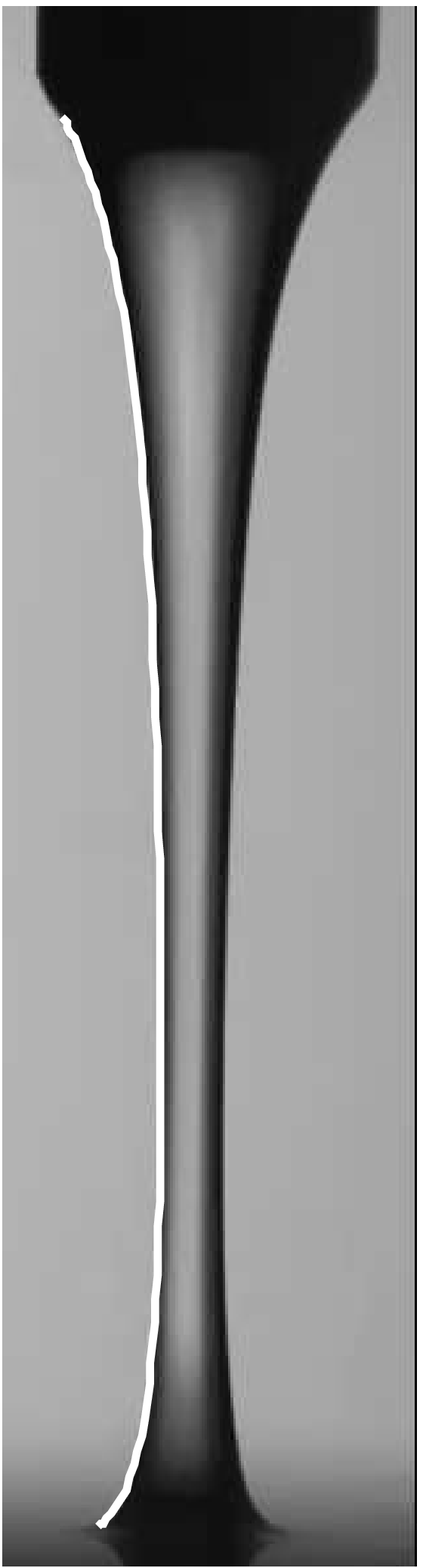}
    \llap{\parbox[b]{80pt}{(\textit{b})\\[215pt]}}
    \caption{(a) Sketch of the flow configuration. (b) Photograph of a steady jet for $\nu=500$ cSt $(\Ka=8.33)$, $R=1.75$ mm $(\Bo=1.38)$, $Q=3.5$ ml$/$min $(\Web=3.05\times 10^{-3})$ and $L=16.6$ mm $(L/R=9.47)$. The white line shows the steady solution of equations~\eqref{eq:cont}--\eqref{eq:curv_adim}.\label{fig:figure1}}
\end{figure}

Figure~\ref{fig:figure1}(a) sketches the configuration under study, where a liquid of density $\rho$, kinematic viscosity $\nu$, and surface tension coefficient $\sigma$ is injected at a constant flow rate $Q$ through a circular tube of radius $R$ whose length is sufficiently large to guarantee a fully-developed velocity profile at its outlet. The resulting free liquid jet is confined in the axial direction through its impact with the free surface of a reservoir of the same liquid placed at a distance $L$ from the injector outlet. To ensure that the results are independent of the downstream boundary condition, we have also performed several experiments where the jet impacts a solid horizontal plate, obtaining nearly identical results. The surrounding gaseous atmosphere, at pressure $p_a$, has a negligible dynamic effect on the jet due to the smallness of the typical liquid velocities.

The experiments were performed in the same setup used by~\citet{Mariano}, with the only addition of a vertical positioning stage located below the injection tube and mounting a platform onto which a reservoir filled with the working liquid was placed. The free surface of the reservoir, located at a distance $L$ from the injector outlet, was impinged by the liquid jet as shown in figure~\ref{fig:figure1}. The liquid was supplied with a Harvard Apparatus PhD Ultra syringe pump, and the free jet was recorded using a Red Lake Motion Pro X high speed camera. To minimise the effect of ambient noise, both the injector and the camera were installed inside a transparent isolation chamber, and the entire system except the syringe pump was placed on an optical table with a passive vibration damping system. Two stainless steel capillary tubes acquired from Tubca were used as injectors, of inner radii $R = 1.5$ mm and $R = 1.75$ mm, that were carefully machined at their tip to ensure that the contact line remained pinned at their inner radii. Two different Newtonian liquids were used in the experiments, both of them PDMS silicon oils from Sigma-Aldrich, whose properties at $25^{\,\circ}$C are $\rho = 970$ kg m$^{-3}$, $\sigma = 21.1$ mN m$^{-1}$, and kinematic viscosities $\nu = 500$ and $\nu = 1000$ mm$^2$ s$^{-1}$, with corresponding values of the Kapitza number, $\Ka = 3 \nu [(\rho^3 g)/\sigma^3]^{1/4}$, of $\Ka = 8.33$ and $\Ka = 16.67$, respectively. In each experimental run, first an injection tube and a silicone oil were selected, and the vertical stage was positioned to provide the desired value of $L$. The syringe pump was programmed to inject an initial liquid flow rate large enough to provide a jetting regime, and to smoothly decrease until the desired target value of $Q$ was achieved.

\subsection{Basic experimental evidence \label{subsec:evidence}}

From the results of the experiments it is deduced that, leaving the presence of coiling apart, there are three possible regimes whose occurrence depends on the values of the control parameters. Let us begin by classifying these regimes following the order in which they are observed when $Q$ decreases smoothly and the rest of the control parameters are fixed:

\begin{itemize}
 
\item[\textbf{1. Steady jetting:}] This regime is illustrated in movie 1 of the supplementary material and in figure~\ref{fig:figure1}(b), which shows a photograph of a steady jet of silicone oil with $\nu=500$ cSt injected through a tube of radius $R=1.75$ mm at a constant flow rate $Q=3.5$ ml$/$min that impinges on the free surface of an oil reservoir placed at a distance $L=16.6$ mm of the injector. The steady jetting regime is stable if $Q$ is larger than a certain critical value, $Q_c$.

\end{itemize}

When $Q<Q_c$ the jet is unstable due to an oscillatory axisymmetric global mode, leading to self-sustained oscillations of shape and velocity whose amplitude increases with time~\citep{SauteryBuggisch,Mariano}. This linear global mode is also referred to as the \emph{breathing mode} throughout the paper. Our experiments reveal that two different nonlinear states may take place if $Q<Q_c$:

\begin{itemize}
 
 \item[\textbf{2. Oscillatory jetting:}] When $Q_b<Q<Q_c$ a limit-cycle state without breakup is achieved, whereby the oscillation amplitude saturates to a certain function of the downstream position (see figure~\ref{fig:figure5}). This regime can be observed in movies 2 and 5 of the supplementary material, as well as in figures~\ref{fig:figure3}--\ref{fig:figure7}. Note that the oscillatory jetting regime is a nonlinear state of the liquid thread, and it should not be confused with the linear breathing mode, which is globally unstable under the conditions where both the oscillatory jetting and dripping regimes take place.
 
 \item[\textbf{3. Dripping:}] When $Q<Q_b$, the amplitude of the oscillations grows until the breakup of the jet takes place, finally leading to a jetting-dripping transition, as illustrated in movies 3 and 4 of the supplementary material, as well as in figure~\ref{fig:figure8}.
 
\end{itemize}

This evidence calls out for an experimental and numerical characterisation of the functions $Q_c\left(\nu,R,L\right)$ and $Q_b\left(\nu,R,L\right)$, that were thus obtained in a wide region of parameter space, as reported in \S\ref{sec:results}.

\subsection{One-dimensional model \label{subsec:model}}

To model the liquid jet we use the dimensionless leading-order one-dimensional mass and momentum equations~\citep{GyC,EggersDupont},
\begin{align}
 \frac{\partial r_j^2}{\partial t}+ \frac{\partial(u r_j^2)}{\partial z} &= 0,\label{eq:cont} \\
 \frac{\partial u}{\partial t}+ u\frac{\partial u}{\partial z} &= 1-\frac{\partial
 \mathcal{C}}{\partial z} + \frac{\Ka}{r_j^2}\frac{\partial}{\partial z}
 \left(r_j^2\frac{\partial u}{\partial z}\right), \label{eq:mom} \\
 \mathcal{C} &= r_j^{-1}\left[1+\left(\displaystyle{\frac{\partial r_j}
 {\partial z}}\right)^2\right]^{-1/2}-\displaystyle{\frac{\partial^2 r_j}{\partial z^2}}
 \left[1+\left(\displaystyle{\frac{\partial r_j}{\partial z}}\right)^2\right]^{-3/2},
 \label{eq:curv_adim}
\end{align}
where the dependent variables $u$ and $r_j$ are the liquid velocity and jet radius respectively, $z$ is the axial coordinate, $t$ is the time and $\mathcal{C}$ is the interfacial curvature. It is worth pointing out that the full expression for the curvature needs to be retained for the model to provide good quantitative predictions of the stability in the case of large injector diameters, for which the jet experiences a strong gravitational stretching close to the neutral conditions and, correspondingly, the stabilizing effect of the axial curvature must be taken into account~\citep{Mariano}. The variables in equations~\eqref{eq:cont}-\eqref{eq:curv_adim} have been made dimensionless using the liquid density $\rho$, the capillary length $l_{\sigma}=(\sigma/\rho\,g)^{1/2}$, and the associated characteristic velocity $\sqrt{g\,l_\sigma}$~\citep{SenchenkoBohr}. Thus, the system~\eqref{eq:cont}-\eqref{eq:curv_adim} only depends on $\Ka$, which is constant for a given liquid and a fixed value of $g$. The solution depends on the constant flow rate, $Q$, and on the injector radius, $R$, through the dimensionless versions of the boundary conditions at $z=0:\,r_j=R/l_\sigma=\Bo^{1/2}$ and $u=U/\sqrt{g\,l_\sigma}=\Web^{1/2}\Bo^{-1/4}$, where $\Bo=\rho g R^2/\sigma$ is the Bond number, $\Web=\rho U^2 R/\sigma$ is the Weber number and $U=Q/(\pi R^2)$ the mean velocity at the nozzle exit. In addition, a Dirichlet boundary condition is imposed for the velocity at the downstream end of the domain, $z=L/l_{\sigma}$: $u=u_{\text{out}}$, as a crude but simple way to represent the impingement of the jet on the reservoir. Specifically, the value of $u_{\text{out}}$ is small enough to properly describe the impact region, since the liquid bath is at rest far from the jet. This simple method provides fairly good results, as illustrated in figure~\ref{fig:figure1}(b) for the particular case of a steady jet. In addition, we have carefully checked that the results barely depend on the value chosen for the outflow velocity, provided that $u_{\text{out}}\lesssim 0.1$. In the case of relatively long jets with $L/R \gtrsim 20$, the value of $u_{\text{out}}$ does not affect the results at all. Indeed, in this limit the impact region is very small compared to the length of the jet, and the outflow boundary condition affects neither the base flow nor its linear and nonlinear stability. In particular, these cases can be easily modelled either by imposing a Neumann outlet boundary condition, or even without imposing any outlet boundary condition at all~\citep{Mariano}. The mathematical model is governed by four control parameters, namely $\Ka$, $\Bo$ and $\Web$ and the dimensionless length of the jet, $L/R=L/l_{\sigma}\,\Bo^{-1/2}$, hereafter scaled with the injector radius for convenience.

\section{Experimental and theoretical results \label{sec:results}}

In this section we present the results of the experiments, as well as the linear stability analysis and the numerical simulations based on the one-dimensional model~\eqref{eq:cont}-\eqref{eq:curv_adim}.
 
\subsection{Linear stability analysis: the role of axial confinement \label{subsec:linear}}

Let us begin by extending the linear stability results of~\citet{Mariano} to account for the effect of $L/R$ on the critical Weber number, $\Web_c$, below which the jet becomes globally unstable. Since the methodology is identical to that presented in~\citet{Mariano}, except for the treatment of the downstream boundary condition, the details of the formulation are provided in Appendix~\ref{app:lsa}. The procedure consists of linearising equations~\eqref{eq:cont}-\eqref{eq:curv_adim} around a given steady basic state according to the following decomposition in temporal normal modes,
\begin{eqnarray}
 r_j (z,t) &=& r_{j0}(z) + \epsilon r_{j1}(z) e^{\omega t},\label{eq:expansionr}\\
 u (z,t) &=& u_{0}(z) + \epsilon u_{1}(z) e^{\omega t},\label{eq:expansionu}
\end{eqnarray}
where $\epsilon \ll 1 $ accounts for the smallness of the perturbation around the base flow $\left[r_{j0}(z),u_0(z)\right]$, $\omega$ is the complex eigenfrequency, and $r_{j1}(z)$ and $u_1(z)$ are the corresponding eigenfunctions. The values of $\omega_r = \Re(\omega)$ and  $\omega_i = \Im(\omega)$ represent the growth rate and the angular frequency of a normal mode, respectively. Accordingly, the jet will be stable if $\max(\omega_r) < 0 $, and unstable if $\max(\omega_r) > 0$. In the latter case, it is important to emphasise that the linearised description given by~\eqref{eq:expansionr}-\eqref{eq:expansionu} is only valid during the initial stages of growth of small disturbances, so that the nonlinear terms are negligible in equations~\eqref{eq:cont}-\eqref{eq:curv_adim}. It is also worth pointing out that the decomposition~\eqref{eq:expansionr}-\eqref{eq:expansionu} does not rely on a local stability analysis, which would involve the usual expansion in normal modes of wavepacket form $e^{\omega t+ ikz}$. Thus, instead of searching for a dispersion relation $D(\omega, k)=0$, here no a priori assumption is made about the shape of the eigenfunctions, whose spatial structure is obtained as part of the solution.

The basic flow, which satisfies the steady version of equations~\eqref{eq:cont}-\eqref{eq:curv_adim}, was calculated using the procedure detailed in Appendix~\ref{app:baseflow}. Once the base flow has been found, the critical conditions for the transition from a globally stable to a globally unstable jet are easily determined as a function of the governing parameters, $\left(\Ka, \Bo, \Web, L/R\right)$, by solving the linear stability problem as explained in Appendix~\ref{app:global}. If the flow is stable, $\max(\omega_r) < 0$, small disturbances are damped and thus a steady flow is expected, like that shown in figure~\ref{fig:figure1}(b) and in movie 1 of the supplementary material. In contrast, if $\max(\omega_r) > 0$ the jet is globally unstable and the development of spontaneous oscillations of increasing amplitude is predicted. In the latter case, two different scenarios emerge according to the experimental evidence described in \S\ref{subsec:evidence}: either the oscillations saturate to a limit cycle without breakup (see figures~\ref{fig:figure3}-\ref{fig:figure7}, and movies 2 and 5 of the supplementary material), or their growth leads to the pinch-off of the liquid column, eventually leading to a dripping regime (see figure~\ref{fig:figure8} and movies 3 and 4 of the supplementary material). Since the oscillation amplitude increases in the downstream direction (see figure~\ref{fig:figure5}) it can be anticipated that the confinement parameter, $L/R$, will have a strong influence not only on the neutral stability curves, but also on the selection of the final nonlinear regime (see \S\ref{subsec:nonlinear}).

\begin{figure}
  \centering
    \includegraphics[width=1.0\textwidth]{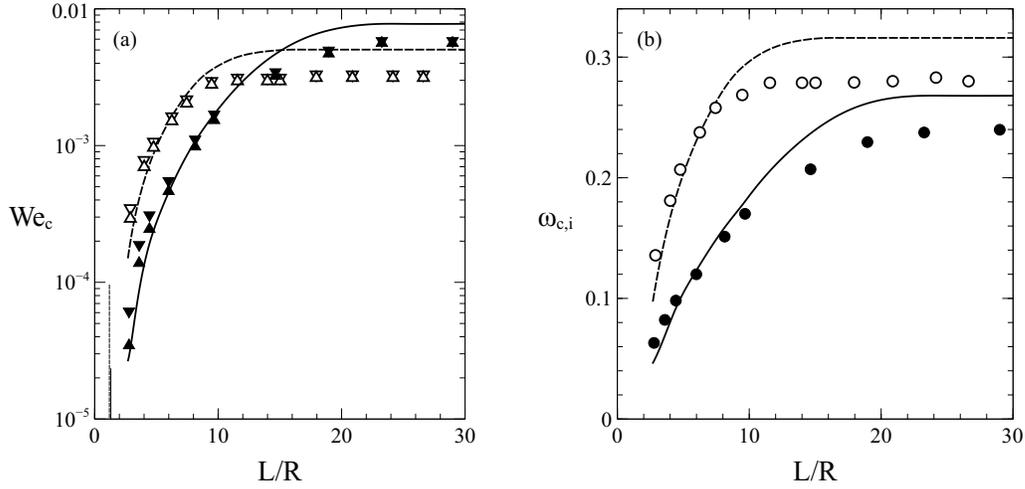}
   \caption{(a) Critical Weber number, $\Web_c$, and (b) corresponding angular frequency, $\omega_{c,i}$, as functions of $L/R$, for $(\Ka,\Bo)=(16.67,1.00)$ (solid lines and filled symbols) and $(\Ka,\Bo)=(8.33,1.38)$ (dashed lines and open symbols). The lines are the neutral curves obtained with the linear stability analysis. The symbols represent the experimental results for the maximum ($\triangledown$) and minimum ($\vartriangle$) values of $\Web_c$, and for the critical frequency ($\circ$). The vertical thin lines in panel (a) are the critical lengths given by the hydrostatic limit, $\Web_c\to 0$ \citep{Kovitz75,Benilov13}.\label{fig:figure2}}
\end{figure}

Both the experiments and the stability analysis reveal that the eigenvalue spectrum is strongly affected by the axial confinement for sufficiently small values of $L/R$. Consequently, $\Web_c$ and $\omega_{c,i}$ are certain functions of $L/R$ which can be easily computed with the methodology described in Appendices~\ref{app:global} and~\ref{app:num_lsa}. The main result is summarized in figure~\ref{fig:figure2}, which shows the good agreement of the experiments (symbols) with the prediction of the linear stability analysis (lines), especially for values of $L/R\lesssim 10$. Figure~\ref{fig:figure2} reveals that both $\Web_c$ and $\omega_{c,i}$ decrease as $L/R$ decreases, indicating that axial confinement stabilises the jet and reduces the critical frequency of the self-sustained oscillations. The symbols $\vartriangle$, $\triangledown$ in figure~\ref{fig:figure2}(a) define the experimentally determined range of neutral stability, due to the fact that the flow rate decreased in smooth ramps programmed with the syringe pump as described in \S\ref{subsec:setup}. For sufficiently long jets, $L/R\gg 1$, the stability properties become independent of $L/R$, reaching the limit already studied by~\citet{SauteryBuggisch} and~\citet{Mariano}. In the opposite limit of strongly confined jets, figure~\ref{fig:figure2}(a) suggests the existence of a vertical asymptote, $\Web_c\to 0$, and a corresponding critical value of the jet length, $L_c/R$. Indeed, the vertical lines plotted in figure~\ref{fig:figure2}(a) are the maximum lengths for which a static axisymmetric liquid bridge between a solid rod and an infinite pool is stable~\citep{Kovitz75,Benilov13}, namely $\left(\Bo,L_c/R\right) \simeq \left(1,1.31\right)$ and $\left(1.38,1.19\right)$. Note from figure~\ref{fig:figure2} that these values of $L_c/R$ are consistent with our results in the hydrostatic limit, $\Web_c\to 0$. It is therefore concluded that axial confinement stabilises the liquid thread, and that the jet length below which such effect is noticeable depends on $\Ka$ and $\Bo$, having values $L/R\lesssim 15$ for $(\Ka,\Bo)=(8.33,1.38)$ and $L/R\lesssim 23$ for $(\Ka,\Bo)=(16.67,1.00)$, as can be deduced from figure~\ref{fig:figure2}. In contrast, the hydrostatic limit reached as $L/R$ decreases is independent of the parameter $\Ka$, which incorporates the liquid viscosity, and is only a function of $\Bo$. Figure~\ref{fig:figure2} also reveals that the quantitative agreement between experiments and theory deteriorates for $L/R\gtrsim 10$, probably due to the limitations of the one-dimensional model. In particular, as emphasised by~\citet{Mariano}, the model does not contemplate the viscous relaxation from the parabolic velocity profile at the injector outlet to the uniform velocity profile achieved downstream. In contrast, the good quantitative agreement found for $L/R\lesssim 10$ may well be due to the fact that the hydrostatic limit, $\Web_c\to 0$, is described exactly by the theoretical model thanks to the use of the full expression for the interfacial curvature in equation~\eqref{eq:curv_adim}.

%%%%%%%%%%%%%%%%%%%%%%%%%%%%%%%%%%%%%%%%%%%%%%%%%%%%%%%%%%%%%%%%%
 
\subsection{Nonlinear stability \label{subsec:nonlinear}}

The present section is devoted to address the influence of the control parameters on the selection of the final jet state under globally unstable conditions, $Q<Q_c$. Although the linear stability analysis presented in \S\ref{subsec:linear} provides values of $Q_c$ in fairly good agreement with experiments, it cannot predict the final state of the jet at large times, which is determined by nonlinear effects. Therefore, in addition to the experiments, we have conducted numerical simulations of equations~\eqref{eq:cont}-\eqref{eq:curv_adim}, which were integrated by means of the simple and efficient method explained in appendices~\ref{app:num_lsa} and~\ref{app:num_nonlinear}. In particular, the latter approach allows to numerically compute the nonlinear saturation process that takes place in the oscillatory jetting regime, and to determine the breakup flow rate, $Q_b$, as a function of the governing parameters $\nu$, $R$ and $L$.

Let us first describe the main characteristics of the oscillatory jetting regime, which takes place for $Q_b<Q<Q_c$ or, in dimensionless terms, for $\Web_b<\Web<\Web_c$. For clarity, hereinafter the values of the variables provided by experiments, linear stability analysis and numerical simulations will be denoted using the superscripts $\left(\,\right)^{\text{exp}}$, $\left(\,\right)^{\text{lsa}}$ and $\left(\,\right)^{\text{num}}$, respectively.

\begin{figure}
     \centering
     \includegraphics[width=0.6\textwidth]{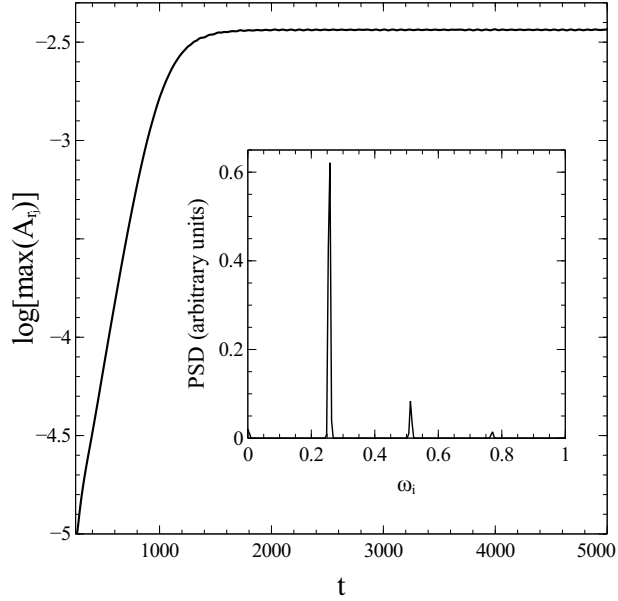}
     \caption{Logarithm of the radius oscillation amplitude at $z=z^*=0.73 L$ as a function of time extracted from a numerical simulation of equations~\eqref{eq:cont}-\eqref{eq:curv_adim} starting from a very small disturbance superimposed on a slightly unstable steady solution, namely $\Ka = 16.67$, $\Bo = 1$, $L/R = 23.26$ and $\Web^{\text{num}}=6.8 \times 10^{-3}<\Web_c^{\text{num}}$. After an initial stage of exponential growth during the linear regime, $t\lesssim 800$, the amplitude saturates to a constant value for $t\gtrsim 2000$ due to nonlinear effects. The inset shows the power spectral density of the saturated oscillations, where the most energetic frequency and its two leading harmonics can be appreciated.\label{fig:figure3}}
\end{figure}

Figure~\ref{fig:figure3} shows a numerical example of the oscillation amplitude growth induced by a very small initial disturbance superimposed on the steady state solution under globally unstable conditions, namely $\Ka=16.67$, $\Bo=1$, $L/R=23.26$ and $\Web^{\text{num}}=6.8\times 10^{-3}<\Web_c^{\text{num}}=\Web_c^{\text{lsa}}=7.88\times 10^{-3}$. In figure~\ref{fig:figure3} the logarithm of the maxima of a pointwise measure of the radius oscillation, $A_{r_j}=r_j(z^*,t)-r_{j0}(z^*)$, where $z^*=0.73 L$, is plotted as a function of time. The choice of the value $z^*=0.73 L$ to measure the jet oscillations was motivated by the fact that their amplitude is sufficiently large at this distance from the injector to allow a precise measurement. In addition, in cases where coiling takes place, like that shown in figure~\ref{fig:figure5}(b), the coiling amplitude at this point is small enough for its influence on the results to be negligible. Three different stages can be clearly identified in figure~\ref{fig:figure3}. First, an initial linear growth regime for $t\lesssim 800$, where the small disturbance increases exponentially with time with a growth rate given by \mbox{$\omega_r^{\text{num}}=3.28\times 10^{-3}$} and $\omega_r^{\text{lsa}}=3.23\times 10^{-3}$ according to the numerical simulation and the linear stability analysis presented in \S\ref{subsec:linear}, respectively. During the second stage, $800\lesssim t \lesssim 2000$, a transition regime takes place where nonlinear effects become important and moderate the exponential growth. Third, for $t\gtrsim 2000$, the oscillation amplitude saturates to a certain constant. The inset of figure~\ref{fig:figure3} represents the power spectral density (PSD) of the saturated signal, clearly showing the dominant angular frequency and its two leading harmonics. The value of the dominant frequency extracted from the numerical PSD is $\omega_i^{\text{num}}=0.259$, to be compared with the corresponding value of the leading eigenmode computed with the linear stability analysis of \S\ref{subsec:linear}, $\omega_i^{\text{lsa}}=0.259$.

\begin{figure}
     \centering
     \includegraphics[width=0.8\textwidth]{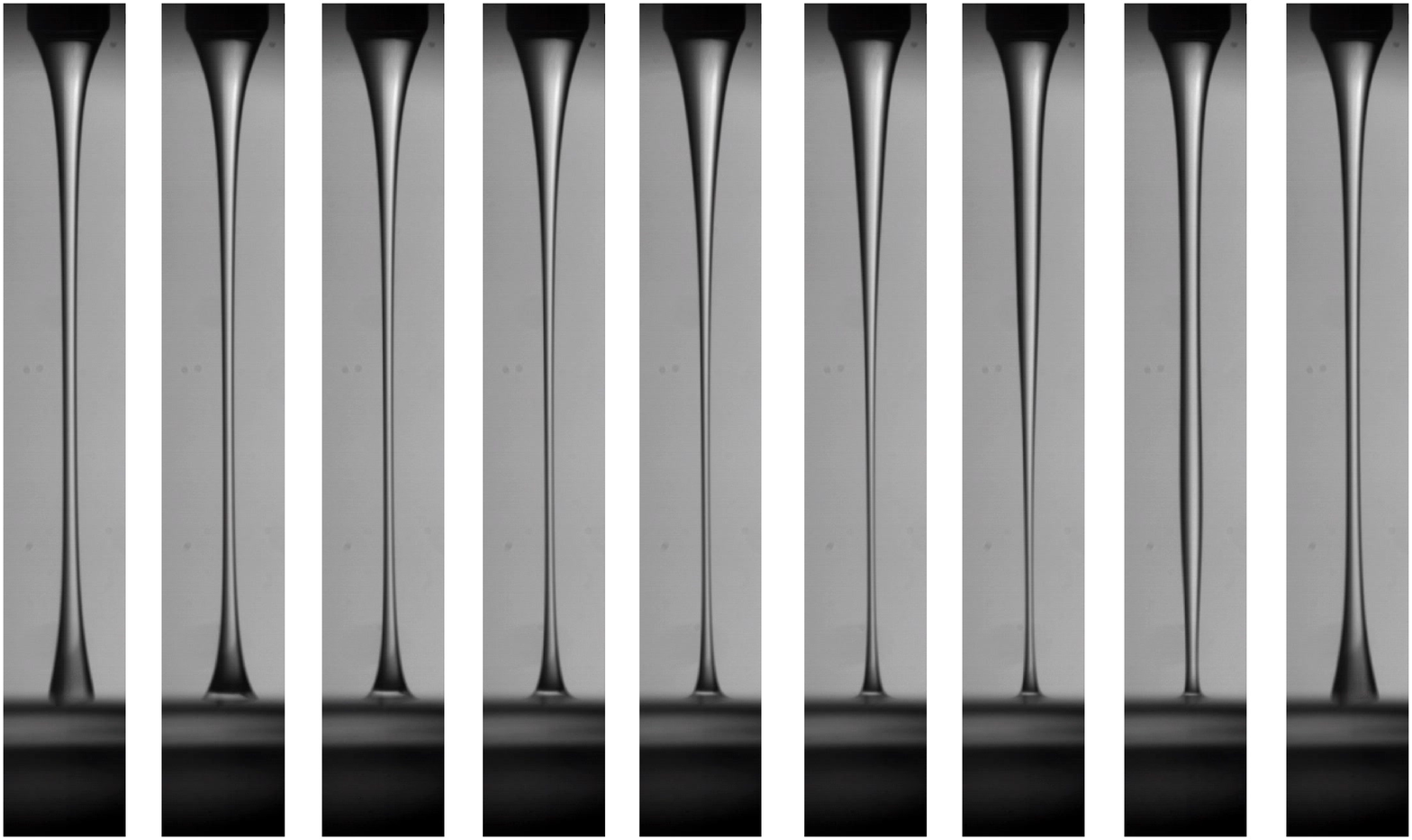}
     \llap{\parbox[b]{325pt}{(\textit{a})\\[165pt]}}
     \\\vspace{2mm}
     \includegraphics[width=0.8\textwidth]{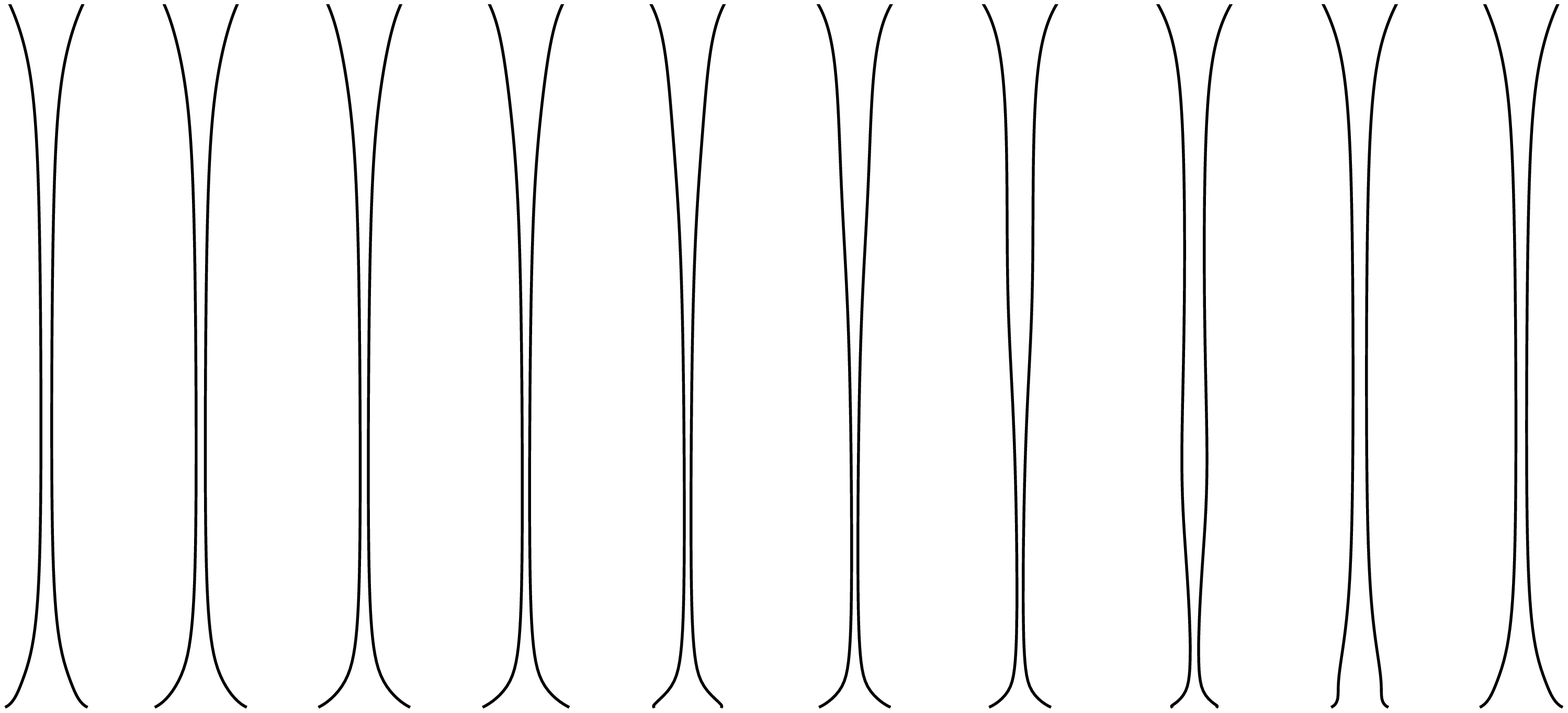}
     \llap{\parbox[b]{325pt}{(\textit{b})\\[122pt]}}
     \caption{(\textit{a}) Photographs of the liquid jet during a period of self-sustained oscillations for $\nu = 500$ cSt ($\Ka = 8.33$), $R = 1.75$ mm ($\Bo = 1.38$), $L = 28$ mm ($L/R = 16$), and $Q^{\text{exp}} = 3.5$ ml$/$min ($\Web^{\text{exp}} = 2.96\times 10^{-3}$), smaller than the experimental critical flow rate, $Q_c^{\text{exp}} = 3.8$ ml$/$min ($\Web_c^{\text{exp}} = 3.49\times 10^{-3}$). The time interval between photographs is $38$ ms, and the oscillation frequency is 3.33 Hz ($\omega_i^{\text{exp}} = 0.258$).
     (\textit{b}) Numerically computed period of self-sustained oscillations for $\Ka = 8.33$, $\Bo = 1.38$, $L/R = 16$ and $\Web^{\text{num}} = 4.43 \times 10^{-3}$, smaller than the critical flow rate provided by the one-dimensional model, $\Web_c^{\text{lsa}} =\Web_c^{\text{num}} = 5.03\times 10^{-3}$. The time interval is $37$ ms and the oscillation frequency is 3.83 Hz ($\omega_i^{\text{num}} = 0.297$). Note that similar distances to the critical point have been chosen in the experiment and in the numerical simulation, namely $\Web^{\text{exp}}_c-\Web^{\text{exp}}\simeq 5.3 \times 10^{-4}$ and $\Web^{\text{num}}_c-\Web^{\text{num}}\simeq 6 \times 10^{-4}$ in panels (a) and (b), respectively.\label{fig:figure4}}
\end{figure}

The typical limit cycle behaviour observed at large times in the oscillatory jetting regime is illustrated in figure~\ref{fig:figure4}. Specifically, figure~\ref{fig:figure4}(a) shows a sequence of photographs captured during one period of self-sustained oscillations, and figure~\ref{fig:figure4}(b) displays the corresponding numerically computed interface profiles under similar conditions. The values of all the dimensionless parameters except $\Web$ are the same in figures~\ref{fig:figure4}(a) and~\ref{fig:figure4}(b), namely $\Ka = 8.33$, $\Bo = 1.38$, $L/R = 16$, for which the critical Weber numbers for the onset of global instability are $\Web_c^{\text{exp}} = 3.49\times 10^{-3}$ and $\Web_c^{\text{lsa}}=\Web_c^{\text{num}} = 5.03\times 10^{-3}$ according to the experiments and to the prediction given by the linear stability analysis and the numerical integration of equations~\eqref{eq:cont}-\eqref{eq:curv_adim}, respectively. Due to the difference in the experimental and numerical values of $\Web_c$, which can also be observed in figure~\ref{fig:figure2}(a), we decided to choose values of $\Web^{\text{exp}}$ and $\Web^{\text{num}}$ in figure~\ref{fig:figure4} such that the distance to the corresponding critical Weber numbers was similar, namely $\Web^{\text{exp}}_c-\Web^{\text{exp}}\simeq 5.3 \times 10^{-4}$ in the experiment of figure~\ref{fig:figure4}(a) and $\Web^{\text{num}}_c-\Web^{\text{num}}\simeq 6 \times 10^{-4}$ in the numerical simulation of figure~\ref{fig:figure4}(b). In view of the results shown in figure~\ref{fig:figure4} it is clear that the simple one-dimensional model used in the present work is able to qualitatively reproduce the main features of the oscillatory jetting regime including not only its frequency, but also the spatial structure of the limit cycle.

\begin{figure}
     \centering
     \includegraphics[width=0.8\textwidth]{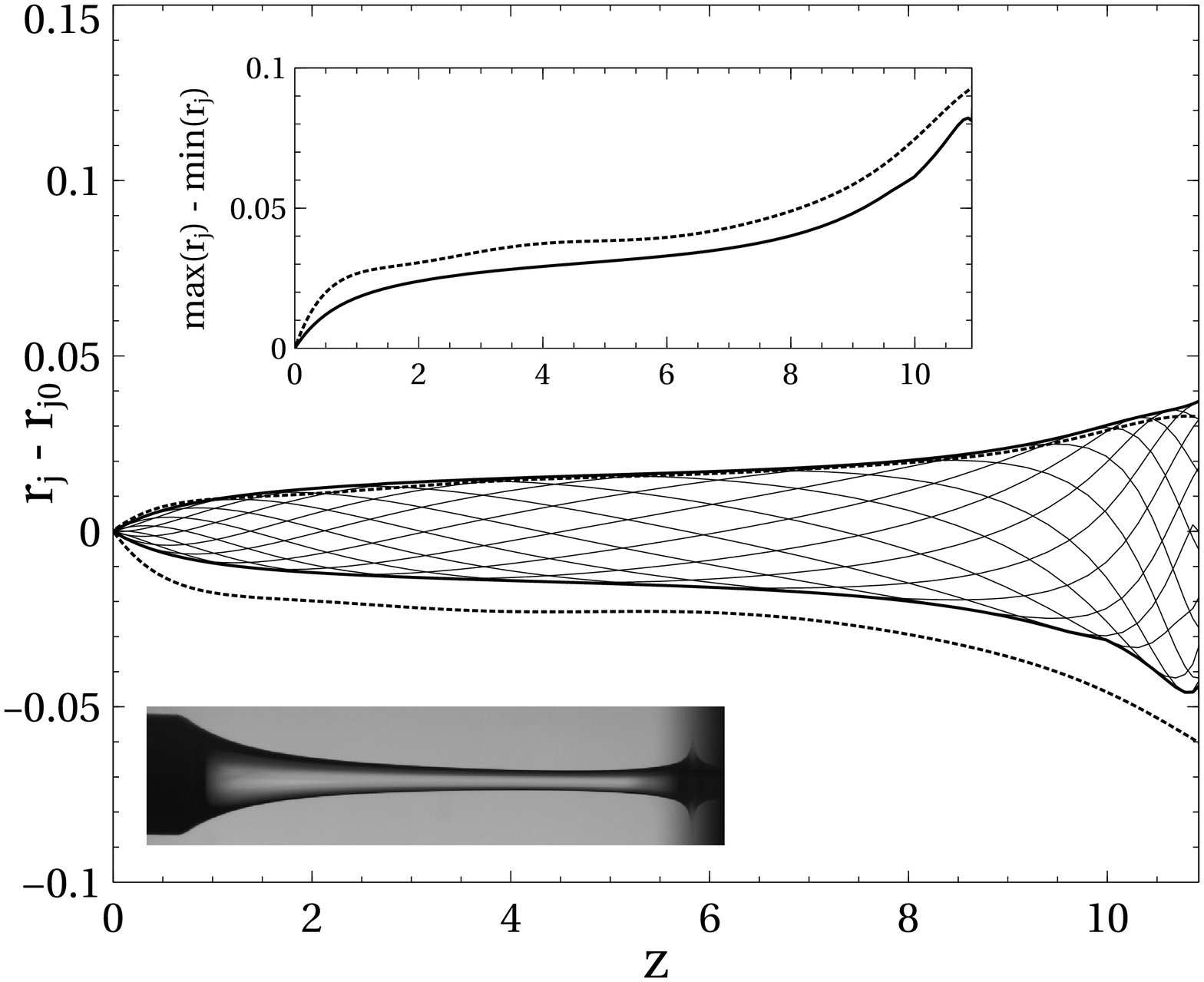}
     \llap{\parbox[b]{320pt}{(\textit{a})\\[240pt]}}
    \hfill
     \includegraphics[width=0.8\textwidth]{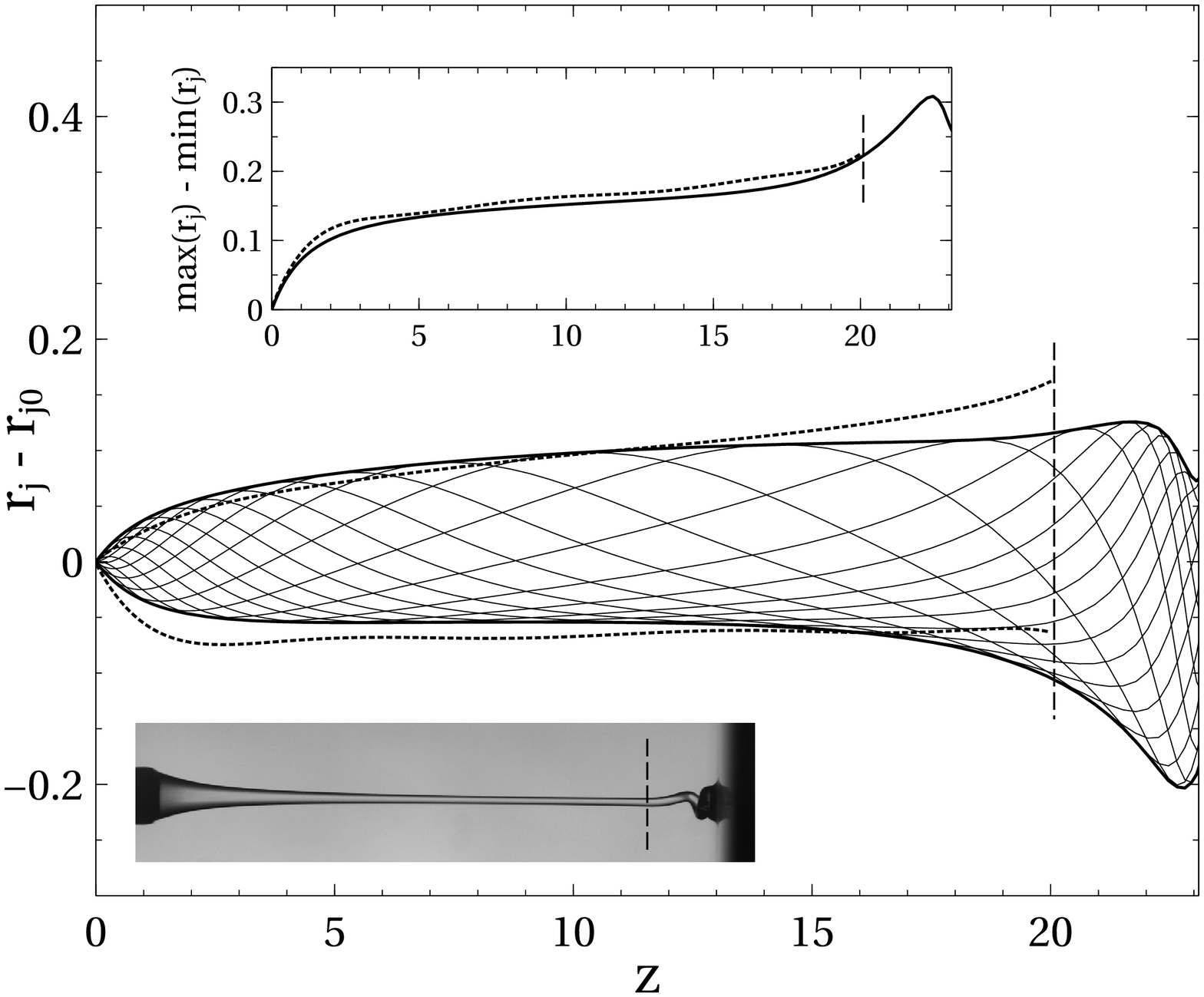}
     \llap{\parbox[b]{320pt}{(\textit{b})\\[240pt]}}
     \caption{Upper and lower envelopes of the oscillations around the basic steady state, $r_j-r_{j0}$, after the saturation to a limit cycle, obtained from experiments (dotted lines) and numerical simulations of equations~\eqref{eq:cont}-\eqref{eq:curv_adim} (solid lines). Numerical jet shapes at several instants during one oscillation period are also plotted as thin solid lines bounded by their corresponding envelope. The inset plots show the difference between the maxima and minima of $r_j$ as a function of $z$. (a) Configuration without coiling, illustated by the inset photograph: $\Ka = 8.33$, $\Bo = 1.38$, $L/R = 9.36$ and values of $\Web<\Web_c$ close to the experimental and numerical neutral conditions, namely $\Web^{\text{exp}} = 2.96\times 10^{-3}$, for which $\omega_{i}^{\text{exp}} = 0.269$, and $\Web^{\text{num}} = 3.75\times 10^{-3}$, for which $\omega_{i}^{\text{num}} = 0.296$. (b) Configuration with coiling, illustated by the inset photograph: $\Ka = 16.67$, $\Bo = 1$, $L/R = 23.26$ and values of $\Web^{\text{exp}} = 3.93\times 10^{-3}$, for which $\omega_{i}^{\text{exp}} = 0.215$, and $\Web^{\text{num}} = 6.418\times 10^{-3}$, for which $\omega_{i}^{\text{num}} = 0.256$. The vertical dashed line marks the upper limit imposed on the value of $z$ in the post-processing of the experiment.\label{fig:figure5}}
\end{figure}

\begin{figure}
     \centering
     \includegraphics[width=0.8\textwidth]{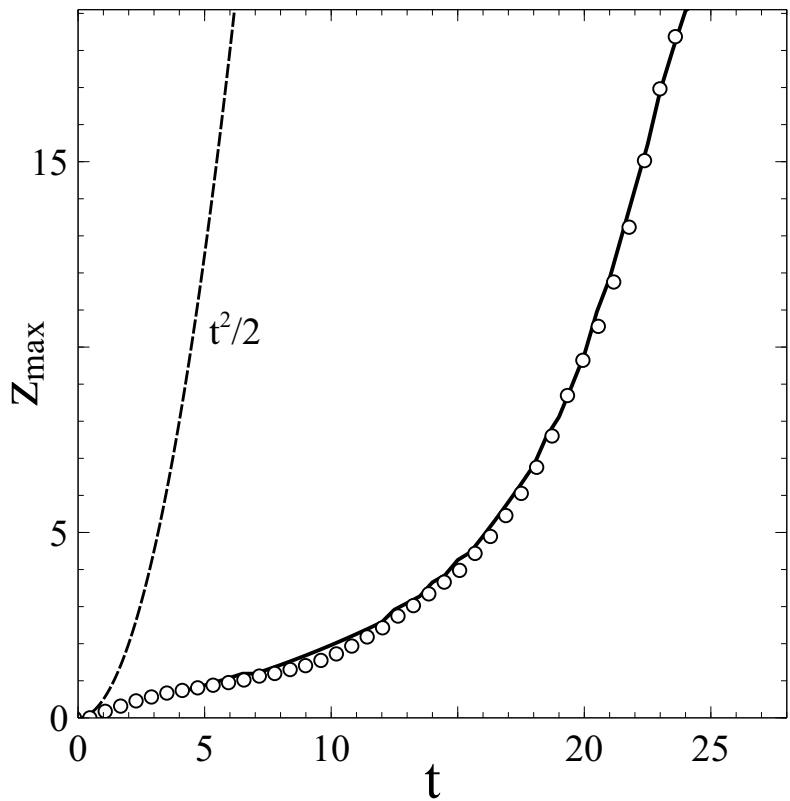}
     \caption{Axial coordinate of the point of maximum radius of the liquid bulge, $z_{\text{max}}$, as a function of time during one cycle of the self-sustained oscillatory jetting state of figure~\ref{fig:figure5}(b), obtained from experiments (symbols) and numerical simulations (solid line). The dashed line shows the free-fall law $t^2/2$. \label{fig:figure6}}
\end{figure}

\begin{figure}
     \centering
     \includegraphics[width=0.8\textwidth]{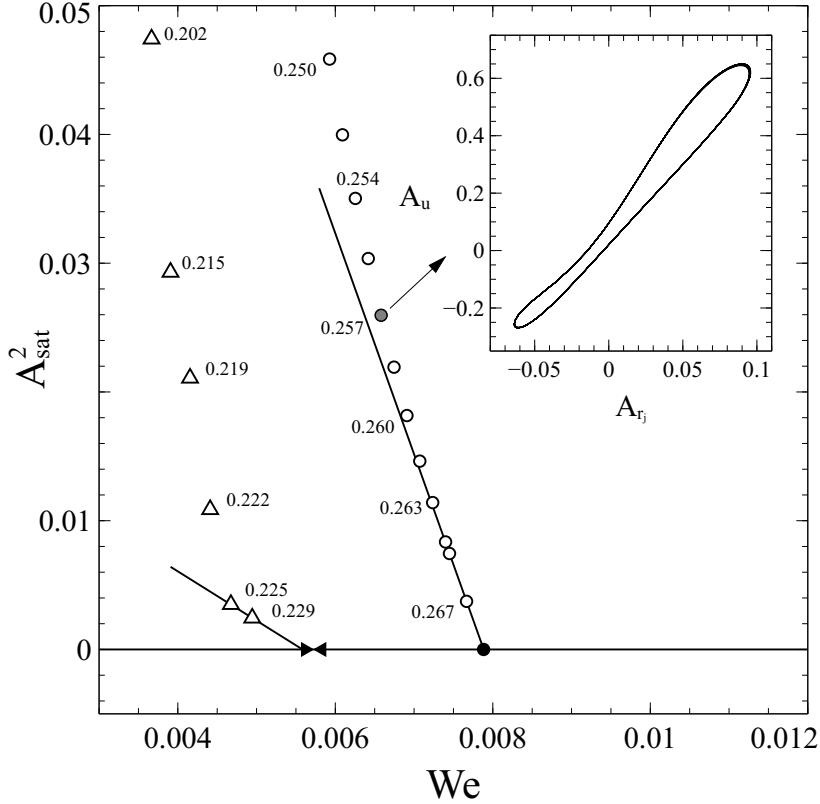}
     \caption{Bifurcation diagram for $\Ka = 16.67$, $\Bo = 1$ and $L/R = 23.26$, where the square of the saturated amplitude of the radius oscillations, $A_{\text{sat}}^2$ (defined in the main text), is represented as a function of $\Web$. Results from experiments ($\vartriangle$) and numerical integrations of equations~\eqref{eq:cont}-\eqref{eq:curv_adim} ($\circ$) are shown. The experimental critical Weber number lies in the range $5.62 \times 10^{-3} < \Web_c^{\text{exp}} < 5.83 \times 10^{-3}$, represented by the symbols $\blacktriangleright$ and $\blacktriangleleft$. The numerical simulations provide the value $\Web_c^{\text{num}} = 7.88 \times 10^{-3}$ ($\bullet$), in agreement with the result of the linear stability analysis, $\Web_c^{\text{lsa}}$. The solid lines are linear fits close to the experimental and numerical neutral points, according to the supercritical Stuart-Landau model. The saturated oscillation frequencies, $\omega_{i}$, are indicated close to several data points. The inset shows the structure of the numerical limit cycle in the $(A_{u},A_{r_j})$ plane associated with the gray-filled circle. The offset between the values of $\Web_c^{\text{exp}}$ and $\Web_c^{\text{num}}$ can also be seen in figures~\ref{fig:figure2} and~\ref{fig:figure9}(d).\label{fig:figure7}}
\end{figure}

A quantitative comparison between the typical experimental and numerical behaviour in the oscillatory jetting regime is provided in figures~\ref{fig:figure5}--\ref{fig:figure7}. In particular, figure~\ref{fig:figure5} shows the numerically computed saturated oscillations of the jet radius around the basic steady state, $r_j\left(z,t\right)-r_{j0}(z)$ as $t \to \infty$ (thin solid lines). The thick solid and dashed lines are, respectively, the upper and lower envelopes of the numerical and experimental oscillations, obtained as the maximum and minimum values of $r_j-r_{j0}$ over time at each value of $z$. The insets display the amplitude of the oscillations as the difference between the upper and the lower envelopes, $\max{\left(r_j\right)}-\min{\left(r_j\right)}$. The oscillation amplitude is seen to increase monotonically downstream until the impact region is reached, explaining why the liquid jets studied herein always break-up close to the free surface of the reservoir, as observed in figure~\ref{fig:figure8} and in movies 3 and 4 of the supplementary material. A particularly interesting aspect of the flow is illustrated in figure~\ref{fig:figure5}(b), which shows that the self-sustained oscillations may coexist with the phenomenon of coiling~\citep{RibeARFM2012}. Indeed, our experiments have revealed new regimes of steady and oscillatory coiling as a function of the parameters of the problem, respectively associated with the steady jetting and with the oscillatory jetting regimes. In the latter case, the coiling is unsteady and its frequency varies enslaved to that of the axisymmetric breathing mode. Another scenario that we have observed is the disappearance of coiling due to the increase of the oscillation amplitude as the value of $\Web$ decreases or the value of $L/R$ increases (see figures~\ref{fig:figure9}a and~\ref{fig:figure9}c). Although these features cannot be predicted using an axisymmetric model, their effect on the saturation amplitude upstream of the impact region is relatively small, as deduced from the results of figure~\ref{fig:figure5}(b). In fact, it is deduced from figure~\ref{fig:figure5} that the quantitative agreement between the experiments and the numerical integration of the one-dimensional model equations~\eqref{eq:cont}-\eqref{eq:curv_adim} is fairly good, provided that the values of $\Web$ are chosen such that $\Web_c-\Web$ have similar values in the experiments and numerical simulations.

Another interesting aspect of the oscillatory jetting regime is the periodic formation of a liquid bulge which falls downstream during each oscillation period, as deduced from figures~\ref{fig:figure4} and~\ref{fig:figure5}, and from movies 2 and 5 of the supplementary material. Figure~\ref{fig:figure6} shows the axial position of the point of maximum radius inside the liquid bulge, $z_{\text{max}}$, as a function of time, under the conditions of figure~\ref{fig:figure5}(b). The symbols and the solid line represent the experimental and numerical results, respectively, while the dashed line is a plot of the free-fall law which, in dimensionless terms, is the function $t^2/2$. The results of figure~\ref{fig:figure6} reveal that the vertical velocity of the bulge is smaller than that associated with free fall during most of its time evolution. Only during the last stages, when the volume accumulated inside the fluid bulge becomes larger than the volume of the liquid filaments connecting the bulge with the injector upstream and with the reservoir downstream, the free-fall law is approached. This fact suggests that the liquid bulge behaves like a freely falling liquid drop when its volume becomes large enough.

The bifurcation diagram represented in Figure~\ref{fig:figure7} shows the squared amplitude of the saturated radius oscillations, $A_{\text{sat}}^2$, as a function of $\Web$, obtained from the \mbox{experiments} ($\vartriangle$) and numerical simulations ($\circ$). Here the saturated amplitude is defined as $A_{\text{sat}}=\max{\left(A_{r_j}\right)}-\min{\left(A_{r_j}\right)}$ with $t>t_{\text{sat}}$, $t_{\text{sat}}$ being a value of time large enough for the asymptotic limit cycle behaviour illustrated by the inset of \mbox{figure~\ref{fig:figure7}} to be reached. Moreover, the amplitude is defined by the same pointwise measure used in figure~\ref{fig:figure3}, namely $A_{r_j}=r_j\left(z^*,t\right)-r_{j0}\left(z^*\right)$ and, similarly, $A_{u}=u\left(z^*,t\right)-u_{0}\left(z^*\right)$. The results shown in figure~\ref{fig:figure7} prove that the axially-confined jets under study behave as hydrodynamic oscillators governed by a supercritical Hopf bifurcation. Indeed, in the case of the numerical simulations, as the value of $\Web^{\text{num}}$ decreases and the critical value $\Web_c^{\text{lsa}}$ given by the linear analysis of figure~\ref{fig:figure2} is crossed ($\bullet$), a branch of finite-amplitude, orbitally stable periodic solutions emerges for $\Web^{\text{num}}<\Web_c^{\text{lsa}}$ ($\circ$). A similar scenario holds also for the experiments, but at smaller values of $\Web^{\text{exp}}$, since $\Web_c^{\text{exp}}<\Web_c^{\text{lsa}}$ as deduced from figure~\ref{fig:figure2}. In addition, figure~\ref{fig:figure7} shows that close to the critical point the amplitude grows as $A_{\text{sat}}^2 \propto \left(\Web_c - \Web\right)$, as expected from the Stuart-Landau equation~\citep{Stuart, Landau1944}. Note also that the frequency decreases as the value of $\Web$ decreases, whilst the growth rate increases leading to shorter saturation times.

\begin{figure}
   \centering
   \includegraphics[width=0.5\textwidth]{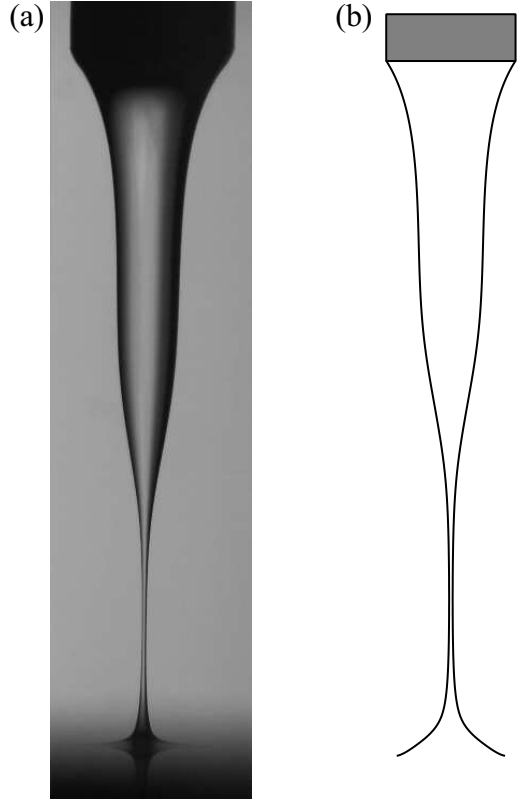}
   \caption{(a) Photograph extracted from movie 3 of the supplementary material, illustrating the incipient breakup of a jet with $\nu = 500$ cSt, $R = 1.75$ mm, $L = 16.6$ mm, injected at a flow rate, $Q = 3.3$ ml$/$min. These conditions are represented in the phase diagrams of figures~\ref{fig:figure9}(a) and~\ref{fig:figure9}(c) with the symbol $\blacktriangle$, which lies within the dripping region. Since $Q<Q_b$, the oscillation amplitude increases and finally leads to dripping through the breakup of a thin thread that connects the meniscus attached to the injector with the liquid reservoir, as clearly observed in movie 3. (b) Instantaneous jet shape at incipient breakup obtained by numerically integrating equations~\eqref{eq:cont}-\eqref{eq:curv_adim} for the equivalent dimensionless configuration, namely $\Ka = 8.33$, $\Bo = 1.38$, $\Web = 2.63 \times 10^{-3}$, $L/R = 9.49$. An animated numerical sequence under these conditions is displayed in movie 4 of the supplementary material.\label{fig:figure8}}
\end{figure}

Since, as revealed by figure~\ref{fig:figure5}, the saturated oscillation amplitude increases with $z$, it can be anticipated that the breakup of the liquid thread will occur as the value of $L/R$ increases, leading to a jetting-dripping transition. Moreover, since the minimum jet radius is always located slightly upstream of the impact region, pinch-off should take place near the reservoir. Indeed, this fact is shown experimentally in the photograph of figure~\ref{fig:figure8}(a), extracted from movie 3 of the supplementary material, and numerically in figure~\ref{fig:figure8}(b), extracted from movie 4 of the supplementary material. Note that the one-dimensional model correctly captures the shape of the interface close to breakup except for the impact region downstream of the minimum radius, probably due to the crude model adopted herein of the outlet boundary condition.

\begin{figure}
     \includegraphics[width=185pt]{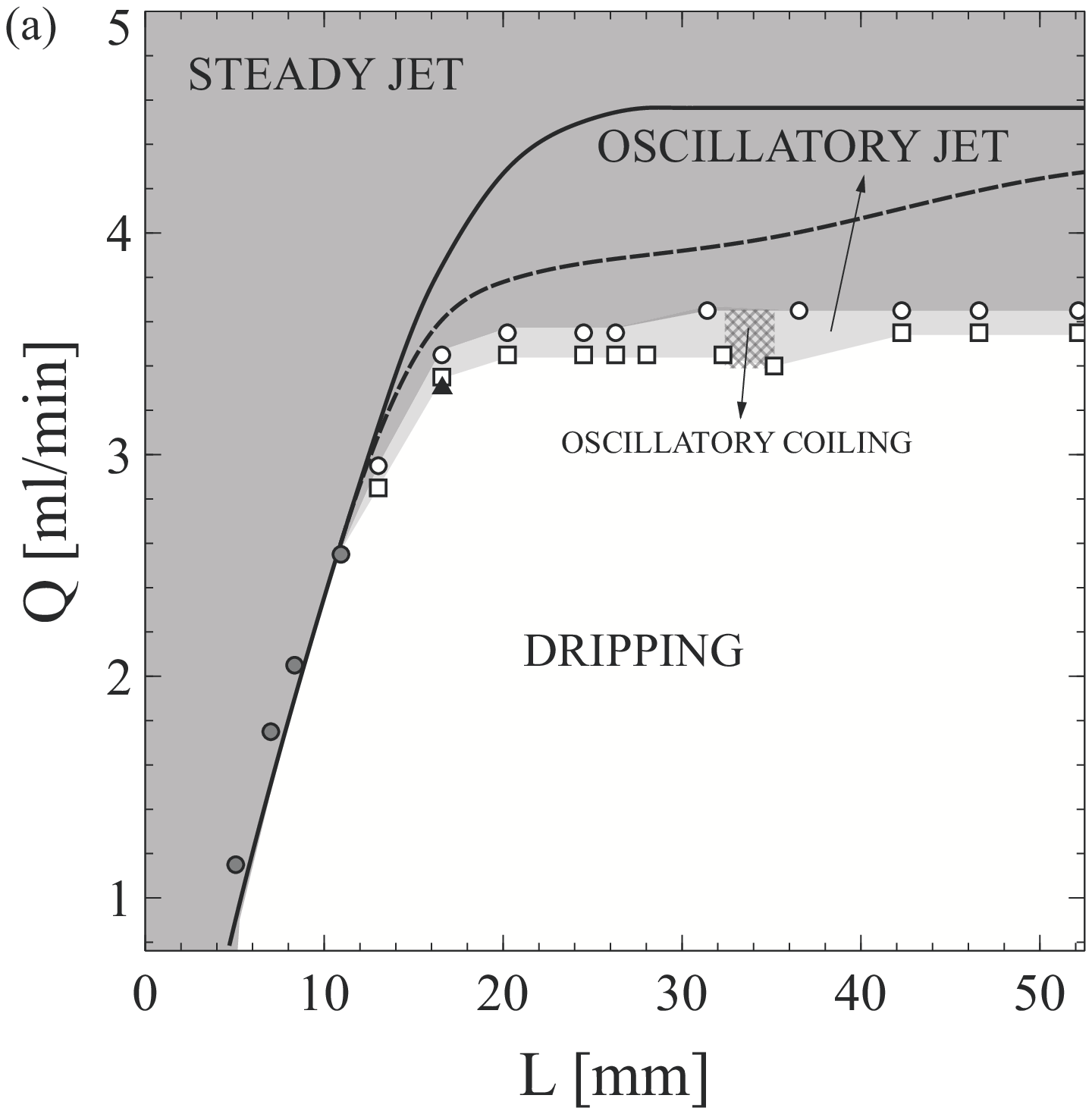}
     \hfill
     \includegraphics[width=185pt]{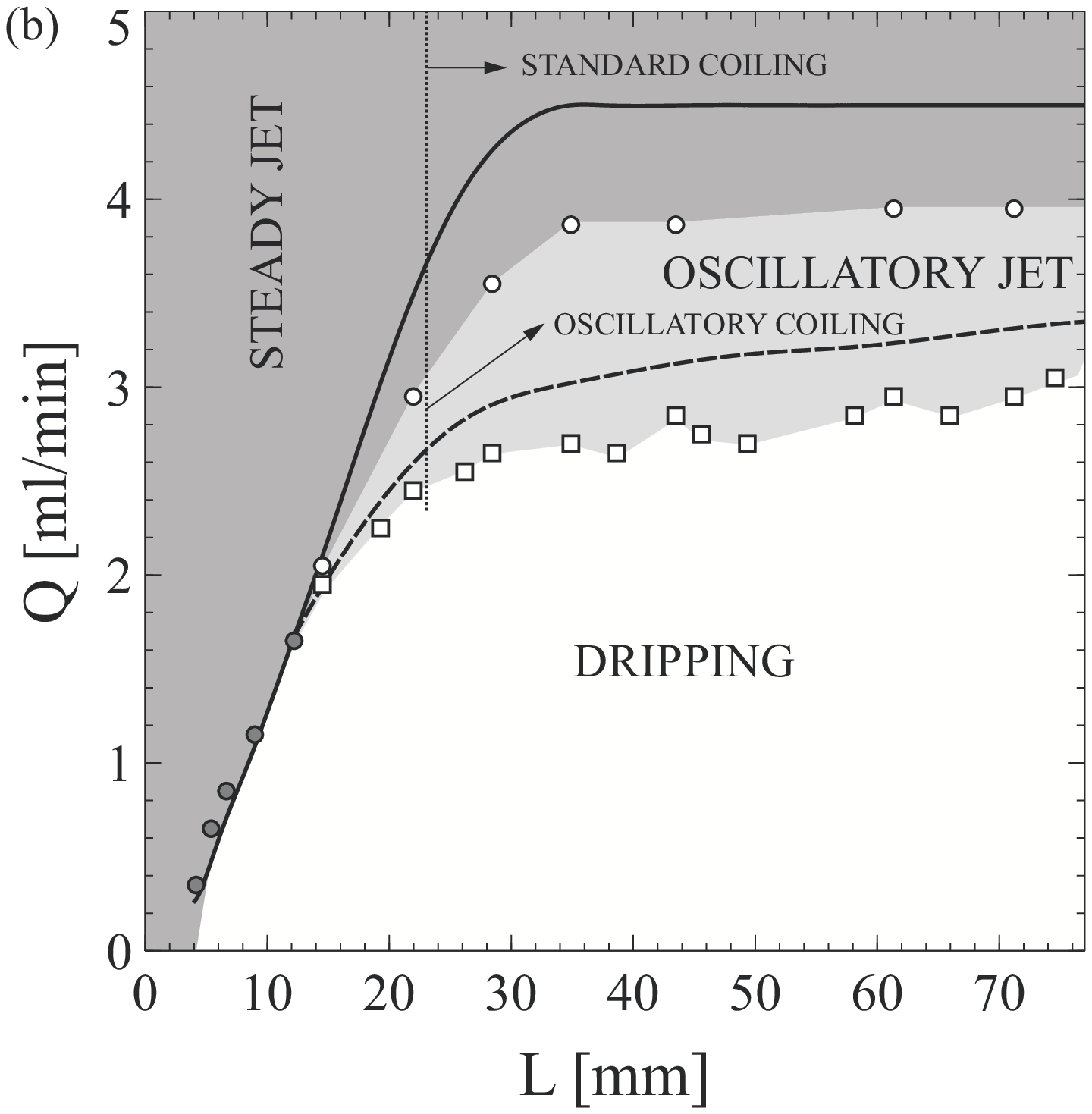}
     \includegraphics[width=185pt]{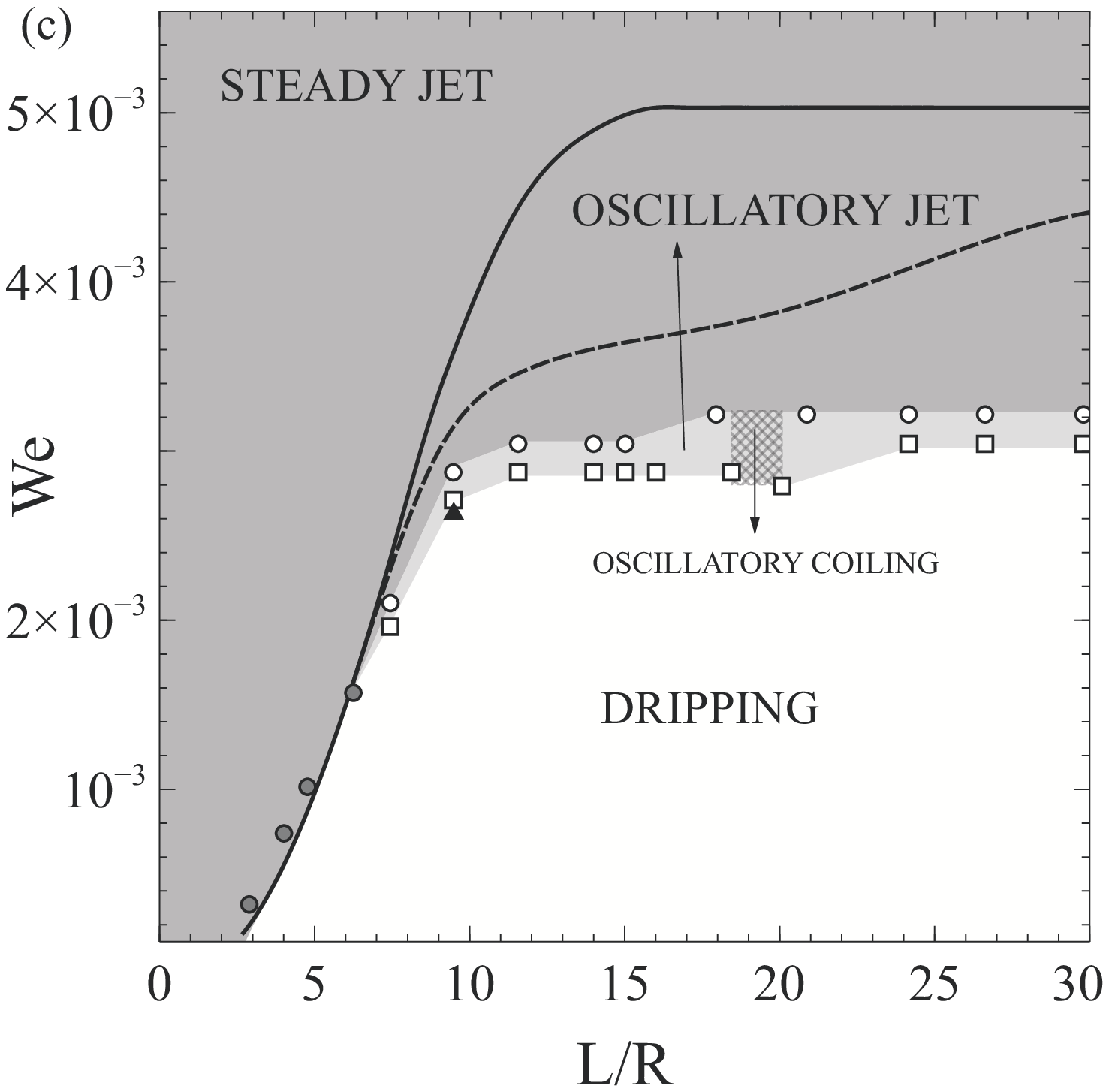}
     \hfill
     \includegraphics[width=185pt]{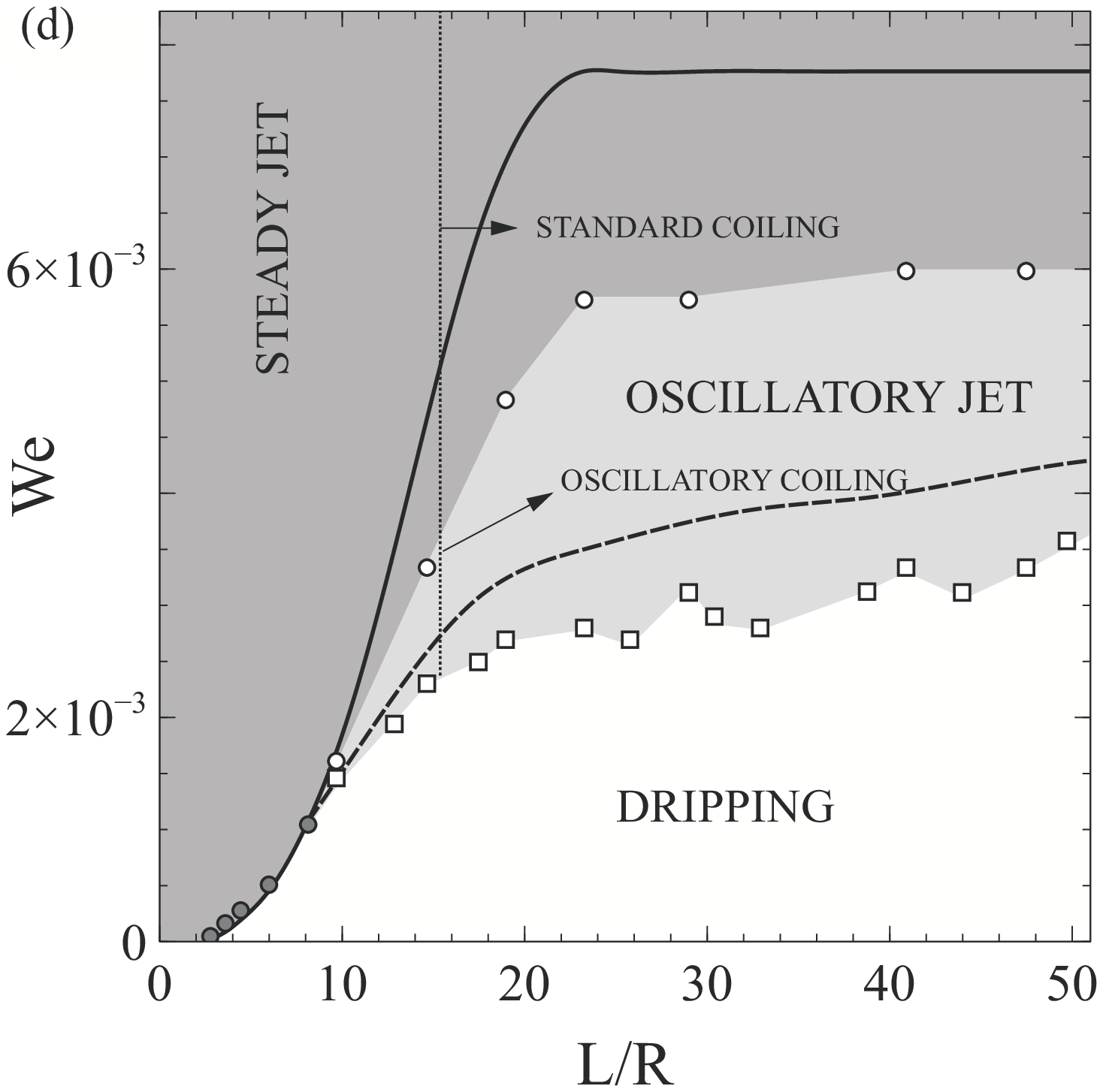}
     \caption{Phase diagrams represented in the dimensional parameter plane $(Q,L)$ (a)-(b), and in the equivalent dimensionless parameter plane $(\Web,L/R)$ (c)-(d), showing the different regimes identified for an axially-confined viscous liquid jet falling under gravity. The dark gray, light gray and unshaded regions correspond with the steady jetting, oscillatory jetting and dripping regimes, respectively. Results are shown for two different combinations of liquid and injector radius, namely (a)-(c) $\nu = 500$ cSt ($\Ka = 8.33$) and $R = 1.75$ mm ($\Bo = 1.38$), and (b)-(d) $\nu = 1000$ cSt ($\Ka = 16.67$), $R = 1.5$ mm ($\Bo = 1$). The mean between the upper and lower limits of the experimental critical flow rate, $Q_c^{\text{exp}}$ ($\Web_c^{\text{exp}}$), and breakup flow rate, $Q_b^{\text{exp}}$ ($\Web_b^{\text{exp}}$), are plotted with the symbols $\circ$ and $\square$, respectively. The solid line is the critical flow rate for the onset of the breathing mode given by the linear stability analysis, $Q_c^{\text{lsa}}$ ($\Web_c^{\text{lsa}}$), and the dashed line represents the function $Q_b^{\text{num}}$ ($\Web_b^{\text{num}}$) obtained from the numerical integration of equations~\eqref{eq:cont}-\eqref{eq:curv_adim}. The gray-filled circles correspond with direct transitions between steady jetting and dripping, while the symbols $\blacktriangle$ represents the conditions of figure~\ref{fig:figure8}. The different coiling states identified in the experiments are indicated.\label{fig:figure9}}
\end{figure}

The preceding discussion naturally leads to the following question: For a constant value of $\Web$, how much can $L/R$ increase while avoiding the breakup of the liquid filament? Or, alternatively, how much can $\Web$ decrease for a constant value of $L/R$ to avoid pinch-off? To address these questions we will make use of the breakup flow rate $Q_b$ or, in dimensionless terms, the breakup Weber number, $\Web_b$. The experimental and numerical phase diagrams represented in figure~\ref{fig:figure9} show the long time regimes reached by the liquid jet in the $(Q,L)$ dimensional parameter plane in panels (a) and (b), and in the equivalent dimensionless parameter plane $(\Web,L/R)$ in panels (c) and (d). Two different combinations of liquid and injector radius are reported in figure~\ref{fig:figure9}, namely $\nu=500$ cSt ($\Ka=8.33$) and $R=1.75$ mm ($\Bo=1.38$) in panels (a) and (c), while $\nu=1000$ cSt ($\Ka=16.67$) and $R=1.5$ mm ($\Bo=1$) in panels (b) and (d). The dark and light gray regions correspond to the steady jetting and oscillatory jetting states according to the experiments, while the dripping regime falls within the unshaded regions. The solid curve is the critical flow rate $Q_c^{\text{lsa}}$ ($\Web_c^{\text{lsa}}$) given by the linear stability analysis of \S\ref{subsec:linear}, and the dashed line is the breakup flow rate $Q_b^{\text{num}}$ ($\Web_b^{\text{num}}$) according to the numerical simulations of equations~\eqref{eq:cont}-\eqref{eq:curv_adim}, obtained by extrapolating the minimum saturated jet radius for several values of $\Web$ slightly larger than $\Web_b^{\text{num}}$. The agreement between $\Web_b^{\text{exp}}$ and $\Web_b^{\text{num}}$ is fairly good for the case with $\Ka=16.67$ ($\nu=1000$ cSt), but quite poor for the case with $\Ka=8.33$ ($\nu=500$ cSt) and values of $L/R\gtrsim 10$. This discrepancy between theory and experiment for values of $L/R\gtrsim 10$ may be due to the limitations of the one-dimensional model that have already been pointed out in the discussion of figure~\ref{fig:figure2}. Indeed, the model does not account for the downstream viscous relaxation of the parabolic velocity profile at the injector outlet. More importantly, our crude modelling of the non-slender impact region where, as shown in figure~\ref{fig:figure8}, the break-up of the thread takes place, may also contribute to the difference between the values of $\Web_b^{\text{exp}}$ and $\Web_b^{\text{num}}$ observed in figure~\ref{fig:figure9}.

A salient feature of the phase maps is the existence of a critical length, $L^*(\nu,R)$ for which $Q_c=Q_b=Q^*$. For values of $L<L^*$ a direct transition between steady jetting and dripping takes place, without the existence of an intermediate oscillating jet state. The corresponding transition points are shown as gray-filled circles in figure~\ref{fig:figure9}, displaying a very good agreement with the value of $Q_c^{\text{lsa}}$. It is important to emphasise that this phenomenon was observed both in the experiments and in the numerical simulations. Let us also point out that, in the cases where a direct transition from steady jetting to dripping takes place, there is no qualitative change in the linear stability mode, which is still the same breathing mode that is destabilised for $Q<Q_c$. Moreover, the nature of the bifurcation that takes place in these cases is also the same Hopf bifurcation shown in figure~\ref{fig:figure7}, the only difference being that the amplitude of the oscillations grows with time without saturation, until break-up takes place, for very small values of $Q_c-Q$. Ideally there should always exist a value of $Q_c-Q$ small enough for the oscillatory jetting regime to be achieved, since the amplitude of the oscillations increases as $(Q_c-Q)^{1/2}$ for sufficiently small values of $Q_c-Q$. In practice, however, the value of $Q_c-Q_b$ is so small in these cases that it cannot be measured in our experiments or numerical simulations.

Figure~\ref{fig:figure9} also reveals that the region of oscillatory jetting decreases as the value of $L/R$ increases, since the amplitude of the oscillations grows with $z$, and thus $Q_b$ increases with $L$, while the value of $Q_c$ reaches an asymptote as $L\to\infty$ (see also figure~\ref{fig:figure2}). Therefore, for values of $L$ larger than those considered herein, it is expected that the values of $Q_c$ and $Q_b$ will intersect, providing a bounded region of oscillatory jetting. Finally, figure~\ref{fig:figure9} shows the important influence of $\nu$, which increases the size of oscillatory jetting region due to its stabilizing role. Similarly, the region of oscillatory jetting becomes smaller as $R$ increases due to the stabilizing effect of axial stretching.

\section{Conclusions \label{sec:conclusions}}

The linear and nonlinear dynamics of jets of viscous liquid falling under gravity and confined in the axial direction has been studied by means of experiments and a simple one-dimensional model~\citep{EggersDupont, GyC} that is extremely efficient from a computational point of view.

The global linear stability analysis reported by~\citet{SauteryBuggisch} and~\citet{Mariano} has been extended in the present work to contemplate the effect of the axial confinement length, $L$. We have shown both experimentally and theoretically that the critical flow rate $Q_c$ for the onset of the axisymmetric breathing mode decreases as $L$ decreases, thus stabilizing the liquid thread. For small enough values of $L$, a limit is reached whereby $\Web_c\to 0$ and the marginal stability point is given by hydrostatics, in agreement with previous works on static liquid bridges~\citep{Kovitz75,Benilov10,Benilov13}.

The nonlinear dynamics has been characterised both experimentally and with numerical integrations of the one-dimensional model. Our results have revealed the existence of a new regime featuring the appearance limit-cycle oscillations of the thread without breakup. The latter oscillatory jetting regime has been thoroughly studied for several combinations of the governing parameters, finally leading to the phase maps presented in figure~\ref{fig:figure9}, that summarises the main conclusions of the present work. In particular, the central role of $L$ is clearly unveiled. In contrast with the linear theory, which only distinguishes between stable and unstable configurations, three distinct long-time nonlinear regimes can be observed in figure~\ref{fig:figure9}: steady jetting, oscillatory jetting and dripping. Although different coiling regimes have also been identified in the experiments, namely a standard and an oscillatory coiling, a detailed study of these coiling regimes is out of the scope of the present work.

The experiments and numerical simulations have revealed that there is a critical value of the confinement length, $L$, below which the oscillatory jetting is no longer observed. This is due to the fact that, as $L$ decreases, $Q_c$ decreases with a larger slope than the critical flow rate associated to the breakup of the jet, $Q_b$. Hence, the size of the oscillatory jetting region decreases as the jet becomes shorter, due to the stabilizing effect of confinement. Consequently, there is a value of $L=L^*$ where $Q_c=Q_b=Q^*$, and for values of $L<L^*$ the transition occurs directly between steady jetting and dripping as $Q$ decreases. The experiments and numerical simulations are in very good agreement in their prediction of the value of $L^*$ and the corresponding $Q^*$.

Future work should contemplate wider ranges of the control parameters, allowing to obtain experimental and numerical phase maps for values of $\Ka$ and $\Bo$ different from those reported herein. Additionally, in view of the new coiling regimes found in our experiments, it could be worthwhile to describe them in detail. Another important extension is the characterization of the physical effects present in typical technological applications, such as the use of liquids with complex rheology or the presence of surfactants. Their inclusion in the analysis is crucial for applications such as 3D printing or additive manufacturing, in which our findings may be of relevance.

\begin{acknowledgments}

The authors thank the Spanish MINECO, Subdirecci\'on General de Gesti\'on de Ayudas a la Investigaci\'on, for its support through projects DPI2014-59292-C3-1-P, DPI2014-59292-C3-3-P and DPI2015-71901-REDT. This research project has been partly financed through European funds.

\end{acknowledgments}

%------------------APPENDIX--------------------%
\begin{appendix}

%%%%%%%%%%%%%%%%%%%%%%% GLSA %%%%%%%%%%%%%%%%%%%%%%%%
\section{Linear stability analysis} \label{app:lsa}

The detailed mathematical formulation of the linear stability problem is provided in the present Appendix.

\subsection{Base flow} \label{app:baseflow}

The base flow [$r_{j0}(z)$, $u_0(z)$] satisfies the steady versions of the system of equations~\eqref{eq:cont}-\eqref{eq:curv_adim}. In particular, the continuity equation~\eqref{eq:cont} allows to substitute the steady state velocity, $u_{0}(z)$, by the base flow jet radius, $r_{j0}(z)$,
\begin{equation}
 \label{eq:mass2}
 u_0 = \frac{q}{r_{j0}^2},
\end{equation}
where $q=\Web^{1/2}\Bo^{3/4}$, is the dimensionless liquid flow rate. In addition, the steady version of the momentum equation~\eqref{eq:mom} reduces to
\begin{equation}
 \label{eq:momentum2}
 u_0 {u_0}' = 1-{\mathcal{C}_0}'+\Ka {r_{j0}}^{-2}({{r_{j0}}^2{u_0}'})',
\end{equation}
where primes indicate derivatives with respect to $z$ and ${\mathcal{C}_0}$ is the mean curvature associated with the steady jet shape, $r_{j0}(z)$, which satisfies equation~\eqref{eq:curv_adim}. Substituting~\eqref{eq:mass2} into~\eqref{eq:momentum2}, the function $r_{j0}(z)$ satisfies
\begin{equation}
 \label{eq:baseflow1}
 -r_{j0}^2{\mathcal{C}_0}'+2q\Ka\left[r_{j0}^{-2}\left(r_{j0}'\right)^2 - r_{j0}^{-1}r_{j0}''\right]+2q^{2}r_{j0}^{-3}r_{j0}'+r_{j0}^2=0,
\end{equation}
where
\begin{equation}
 \label{eq:baseflow2}
 -r_{j0}^2{\mathcal{C}_0}' = \frac{r_{j0}'}{\left[1+\left(r_{j0}'\right)^2\right]^{1/2}}+\frac{r_{j0}r_{j0}'r_{j0}''+r_{j0}^2r_{j0}'''}{\left[1+\left(r_{j0}'\right)^2\right]^{3/2}}-\frac{3r_{j0}^2r_{j0}'\left(r_{j0}''\right)^2}{\left[1+\left(r_{j0}'\right)^2\right]^{5/2}},
\end{equation}
to be solved with the boundary conditions
\begin{eqnarray}
 z = 0&:& r_{j0} = \Bo^{1/2},\label{eq:bc_z0_rj0}\\
 z = 0&:& u_0 = \Web^{1/2}\Bo^{-1/4},\label{eq:bc_z0_u0}\\
 z=\frac{L}{l_\sigma}&:& u_0 = u_{\text{out}},\label{eq:bc_lj_u0}
\end{eqnarray}
as discussed in \S\ref{subsec:model}. Note that equations~\eqref{eq:baseflow1}-\eqref{eq:baseflow2} are the same as those solved in ~\citet{Mariano} (hereinafter R13), the only difference being the outlet boundary condition~\eqref{eq:bc_lj_u0}, which in the present work is imposed as a Dirichlet condition, while a free boundary condition was considered in~R13. The numerical method used herein to solve equations~\eqref{eq:baseflow1}-\eqref{eq:baseflow2} is also the same as that used by~R13, and is explained in Appendix~\ref{app:num_lsa}.

\subsection{Linear stability problem} \label{app:global}

Substituting~\eqref{eq:expansionr}-\eqref{eq:expansionu} into~\eqref{eq:cont}-\eqref{eq:curv_adim}, the $O(\epsilon)$ eigenvalue problem can be written in the following compact form,
\begin{equation} \label{eq:eigen}
  \left[
   \begin{array}{cc}
     \mathcal{M}^c_{r_j} & \mathcal{M}^c_u\\
     \mathcal{M}^m_{r_j} & \mathcal{M}^m_u
   \end{array}
  \right]\left[
   \begin{array}{c}
     r_{j1}\\
     u_1
   \end{array}
   \right]=\omega\left[
   \begin{array}{c}
     r_{j1}\\
     u_1
   \end{array}\right],
 \end{equation}
 with $\mathcal{M}_i^j$ denoting the following differential operators,
 \begin{align}
 \mathcal{M}_{r_j}^c &= -\frac{q}{r_{j0}^2}\,{\rm D}+
 \frac{q\,r'_{j0}}{r_{j0}^3}\,{\rm I}\,,\label{eq:App0}\\
 \mathcal{M}_u^c &= -\frac{r_{j0}}{2}\,{\rm D} - r'_{j0}\,{\rm I}\,,\\
 \mathcal{M}_{r_j}^m &= -
 4\Ka q\left(\frac{r'_{j0}}{r_{j0}^4}\,{\rm D}-\frac{\left(r'_{j0}\right)^2}
 {r_{j0}^5}\,{\rm I}\right) + \sum_{k=1}^4 \mathcal{S}^{2k-1}\,\mathcal{H}_k \,,\\
 \mathcal{M}_u^m &= \Ka\left({\rm D}^2+\frac{2r'_{j0}}{r_{j0}}\,{\rm D}\right)
 -\frac{q}{r_{j0}^2}\,{\rm D}+\frac{2q\,r'_{j0}}{r_{j0}^3}\,{\rm I}\,.\label{eq:kk}
 \end{align}
 
In~\eqref{eq:App0}-\eqref{eq:kk}, $\rm{I}$ is the identity,
${\rm D}^n \equiv {\rm d}^n/{\rm d} z^n$, $\mathcal{S}(z)=[1+(r'_{j0})^2]^{-1/2}$ and
\begin{align}
  \mathcal{H}_1 &= \frac{1}{r_{j0}^2}\,{\rm D} - \frac{2r'_{j0}}{r_{j0}^3}\,{\rm I},\\
  \mathcal{H}_2 &= {\rm D}^3 + \frac{r'_{j0}}{r_{j0}}\,{\rm D}^2 -
  \left[\frac{\left(r'_{j0}\right)^2}{r_{j0}^2} - \frac{r''_{j0}}{r_{j0}}\right]{\rm D} -
  \frac{r'_{j0}\,r''_{j0}}{r_{j0}^2}\,{\rm I},\\
  \mathcal{H}_3 &= -6\,r'_{j0}\,r''_{j0}\,{\rm D}^2 -
  3\left[\frac{\left(r'_{j0}\right)^2\,r''_{j0}}{r_{j0}} + \left(r''_{j0}\right)^2 +
  r'_{j0}\,r'''_{j0}\right]{\rm D},\\
  \mathcal{H}_4 &= 15 \left(r''_{j0}\right)^2\,\left(r'_{j0}\right)^2\,{\rm D}.\label{eq:Append}
\end{align}
Note that equation~\eqref{eq:eigen} is a linear eigenvalue problem, complemented with the boundary conditions
\begin{eqnarray}
 z = 0&:& u_1 = 0,\label{eq:bc_z0_u1}\\
 z = 0&:& r_{j1} = 0,\label{eq:bc_z0_rj1}\\
 z = \frac{L}{l_\sigma}&:& u_1 = 0.\label{eq:bc_lj_u1}
\end{eqnarray}
Equations~\eqref{eq:bc_z0_u1} and~\eqref{eq:bc_z0_rj1} represent the pinned contact line and constant flow rate conditions, respectively. However, equation~\eqref{eq:bc_lj_u1} just represents in a crude way the fact that the disturbed jet velocity is small in the impact region onto the bath at rest. If a small enough value of $u_{\text{out}}$ is assumed for the base flow, the results of the stability analysis do not vary significantly, thereby justifying the use of equation~\eqref{eq:bc_lj_u1}. Moreover, for values of the jet length $L/R \gtrsim 20$ the results are independent of the value of $u_{\text{out}}$, and in fact we have checked that both the base flow and its linear stability are virtually the same whether imposing a Neumann condition at $z=L/l_{\sigma}$, or even not imposing any condition at all, in agreement with the results of~R13.

%%%%%%%%%%%%%%%%%%%%%%% NUMERICAL METHODS %%%%%%%%%%%%%%%%%%%%%%%%
\section{Numerical methods} \label{app:num}

In the present Appendix we describe the numerical methods used for the computation of the steady jet and its linear stability, as well as the integration of equations~\eqref{eq:cont}-\eqref{eq:curv_adim}.

\subsection{Base flow and linear stability analysis} \label{app:num_lsa}

To solve both equation~\eqref{eq:baseflow1} for $r_{j0}$, and the system~\eqref{eq:eigen} for the eigenvalues $\omega$ and the corresponding eigenfunctions, we discretized the differential operators using a Chebyshev collocation method~\citep{Canuto}. To that end, the physical domain, $0 \leq z \leq L/l_{\sigma}$ is mapped into the interval $-1\leq y\leq 1$ by means of the transformation 
\begin{equation}
z = \frac{b L/l_{\sigma} (1+y)}{2b+ L/l_{\sigma}(1-y)}, 
\end{equation}
where $b$ is a parameter that controls the clustering of nodes at $z=0$ and $z=L/l_{\sigma}$. Derivatives with respect to $z$ are calculated using the standard Chebyshev differentiation matrices and the chain rule. The nonlinear differential equation~\eqref{eq:baseflow1} with boundary conditions~\eqref{eq:bc_z0_rj0}-\eqref{eq:bc_lj_u0} is solved first using an iterative Newton-Raphson method. Once $r_{j0}$ is known at the $N$ Chebyshev collocation points, the discretized version of~\eqref{eq:eigen}, which results in a linear algebraic eigenvalue problem to determine the $2N$ eigenvalues $\omega^k$, $k=1\ldots 2N$, and their corresponding eigenfunctions $(r_{j1}^k,u_1^k)$ at the $N$ collocation points, is solved using standard Matlab routines. Notice that the eigenfunctions must satisfy the boundary conditions~\eqref{eq:bc_z0_u1}-\eqref{eq:bc_lj_u1}. Although all the results reported in the present paper were computed with values of $N$ and $b$ within the ranges $30\leq N\leq 200$ and $5\leq b\leq 80$, we have carefully checked that the leading eigenvalues and eigenfunctions are insensitive to the values of these parameters.

%%%%%%%%%%%%%%%%%%%%%%%%%%%%%%%%%%%%%%%%%%%%%%%%%%%%%%%%%%%%%%%%
\subsection{Direct numerical simulation} \label{app:num_nonlinear}

The system of equations~\eqref{eq:cont}-\eqref{eq:curv_adim}, supplemented with the boundary conditions discussed above were solved with a very simple and efficient method of lines in which the spatial derivatives were approximated using the Chebyshev collocation method described in \S\ref{app:num_lsa}. The resulting system of $2N$ nonlinear ordinary differential equations were integrated in time with an initial condition taken as the base flow $[r_{j0}(z),u_0(z)]$, slightly perturbed by a Gaussian function of very small amplitude. The \textsc{ode23t} routine from the Matlab ODE suite was chosen, since Dirichlet or Neumann boundary conditions can be easily imposed through a mass matrix, and also allows to implement the Jacobian matrix of the nonlinear system to improve the speed and accuracy of the computations. The solver \textsc{ode23t} uses the Bogacki--Shampine algorithm, which is a method based on two single-step formulas, of second and third order respectively, and computes three stages for each time step. Briefly stated, to calculate the temporal evolution of the $r_j$ and $u$, the semi-discretised equations were written in the matrix form
\begin{equation}
\frac{\D \bm{y}}{\D t} = \mathsfbi{B}(\bm{y},t),
\end{equation}
where the column vector $\bm{y} = [\bm{r}_j(\bm{z},t), \bm{u}(\bm{z},t)]^{\text{T}}$ contains the values of the jet radius and axial velocity computed at the $N$ Chebyshev collocation points, $\bm{z}$, and the matrix $\mathsfbi{B}$ is
\begin{equation}
\mathsfbi{B}(\bm{y},t) =   \left[\begin{array}{cc}
 \mathsfbi{B}^{c}_{r_j}   &  \mathsfbi{B}^{c}_u \\
 \mathsfbi{B}^{m}_{r_j}   &  \mathsfbi{B}^{m}_u \\
\end{array}\right]_{2N \times 2N} \bcdot \left[\begin{array}{c} \mathbf{r}_{j} (\mathbf{z} = 0,t)\\ \vdots \\ \mathbf{r}_{j} (\mathbf{z} = L/l_{\sigma},t) \\  \mathbf{u} (\mathbf{z} = 0,t)\\ \vdots \\  \mathbf{u} (\mathbf{z} = L/l_{\sigma},t) \end{array}\right]_{2N \times 1} + \left[\begin{array}{c} 0 \\ \vdots \\ 0 \\ 1 \\ \vdots \\ 1 \end{array}\right]_{2N \times 1}. \label{eq:matrixB}
\end{equation}
The $N \times N$ submatrices $\mathsfbi{B}_i^j$ appearing in equation~\eqref{eq:matrixB} are
\begin{equation}
\mathsfbi{B}^{c}_{r_j} = -\text{diag}\left(\bm{u}\right) \bcdot {\mathsfbi{D}}, \label{eq:simp1}
\end{equation}
\begin{equation}
\mathsfbi{B}^{c}_u = -\frac{1}{2}\, \text{diag}\left(\bm{r}_j\right) \bcdot {\mathsfbi{D}},
\end{equation}
\begin{equation}
\mathsfbi{B}^{m}_{r_j} = \sum _{n = 1}^{3} \mathsfbi{C}_n \bcdot {\mathsfbi{D}}^n,
\end{equation}
\begin{equation}
\mathsfbi{B}^{m}_u = -\text{diag}\left(\bm{u}\right) \bcdot \mathsfbi{D} + \Ka \left[\mathsfbi{D}^2 + 2\,\text{diag}\left(\bm{r}_j\right)^{-1} \bcdot \text{diag}\left(\mathsfbi{D} \bcdot \bm{r}_j\right)\bcdot \mathsfbi{D} \right],
\end{equation}
where $\mathsfbi{D}^n$ is the $N\times N$ $n$-th order Chebyshev differentiation matrix, $\text{diag}\left(\cdot\right)$ maps an $N$-tuple to the corresponding diagonal matrix, and the curvature gradient diagonal matrices, $\mathsfbi{C}_n$, are
\begin{equation}
\mathsfbi{C}_1 = \text{diag}\left(\bm{r}_j\right)^{-2} \bcdot \left[ \mathsfbi{I} + \text{diag}\left(\mathsfbi{D} \bcdot \bm{r}_j\right)^{2}  \right]^{-1/2} -3 \,\text{diag}\left(\mathsfbi{D}^2 \bcdot  \bm{r}_j\right)^{2} \bcdot \left[ \mathsfbi{I} + \text{diag}\left(\mathsfbi{D} \bcdot \bm{r}_j\right)^{2}  \right]^{-5/2},
\end{equation}
\begin{equation}
\mathsfbi{C}_2 = \text{diag}\left(\mathsfbi{D} \bcdot \bm{r}_j\right) \bcdot \text{diag}\left( \bm{r}_j\right)^{-1} \bcdot \left[ \mathsfbi{I} + \text{diag}\left( \mathsfbi{D} \bcdot \bm{r}_j\right)^{2}  \right]^{-3/2},
\end{equation}
\begin{equation}
\mathsfbi{C}_3 =  \left[\mathsfbi{I} + \text{diag}\left(\mathsfbi{D} \bcdot \bm{r}_j\right)^{2}  \right]^{-3/2},\label{eq:simpN}
\end{equation}
where $\mathsfbi{I}$ is the $N\times N$ identity matrix. Note that, for clarity, in equations~\eqref{eq:simp1}-\eqref{eq:simpN} the functional dependence of $\bm{r}_j$ and $\bm{u}$ on $(\bm{z},t)$ has been omitted.

The boundary conditions~\eqref{eq:bc_z0_rj0}--\eqref{eq:bc_lj_u0} described in~\S\ref{subsec:model} can be readily implemented by imposing
\begin{equation}
\frac{\D \boldm{r}_j}{\D t}(\boldm{z} = 0,t) = \frac{\D \boldm{u}}{\D t}(\boldm{z} = 0,t)= \frac{\D \boldm{u}}{\D t}(\boldm{z} = L/l_{\sigma},t) = 0.
\end{equation}

\end{appendix}

%--------------------------------------BIBLIOGRAPHY---------------------------------------%
\bibliographystyle{jfm}
%\bibliography{free_jet}

\end{document}